\documentclass[12pt,preprint]{aastex}
\usepackage{amsmath}

\usepackage{hyperref}

\newcommand{\eb}{\begin{equation}}
\newcommand{\ee}{\end{equation}}
\newcommand{\PS}{Pan-STARRS~}

\shorttitle{Block Adjustment for Pan-STARRS}
\shortauthors{Berghea et al.}

\begin{document}

\title{A Global Astrometric Solution for Pan-STARRS referenced to ICRF2 }

\author{C.~T.~Berghea\altaffilmark{1}, V.~V.~Makarov\altaffilmark{1}, J.~Frouard\altaffilmark{1}, G.~S.~Hennessy\altaffilmark{1}, B.~N.~Dorland\altaffilmark{1}, D.~R.~Veillette\altaffilmark{1}, R.~P.~Dudik\altaffilmark{1}, E.~A.~Magnier\altaffilmark{2}, W. S. Burgett\altaffilmark{3}, K. C. Chambers\altaffilmark{2}, L. Denneau\altaffilmark{2}, H. Flewelling\altaffilmark{2}, N. Kaiser\altaffilmark{2}, J. L. Tonry\altaffilmark{2}, R. J. Wainscoat\altaffilmark{2}, B.~Sesar\altaffilmark{4}}

\altaffiltext{1}{U.S. Naval Observatory (USNO), 3450 Massachusetts Avenue NW, Washington, DC 20392, USA}
\altaffiltext{2}{Institute for Astronomy, University of Hawaii at Manoa, Honolulu, HI 96822, USA}
\altaffiltext{3}{GMTO Corp., 465 N. Halstead St., Suite 250, Pasadena, CA   91107}
\altaffiltext{4}{Max Planck Institute for Astronomy, K\"{o}nigstuhl 17, D-69117 Heidelberg, Germany}

\begin{abstract}

We describe development and application of a {\it Global Astrometric Solution} (GAS) to the problem of Pan-STARRS1 (PS1) astrometry.  Current PS1 astrometry is based on differential astrometric measurements using 2MASS reference stars, thus PS1 astrometry inherits the errors of the 2MASS catalog.  The GAS, based on a single, least squares adjustment to approximately 750k ``grid stars'' using over 3000 extragalactic objects as reference objects, avoids this catalog-to-catalog propagation of errors to a great extent.  The GAS uses a relatively small number of Quasi-Stellar Objects (QSOs, or distant AGN) with very accurate ($<$1 mas) radio positions, referenced to the ICRF2. These QSOs provide a hard constraint in the global least squares adjustment. Solving such a system provides {\it absolute astrometry} for all the stars simultaneously. The concept is much cleaner than conventional astrometry but is not easy to perform for large catalogs. In this paper we describe our method and its application to Pan-STARRS1 data. We show that large-scale systematic errors are easily corrected but our solution residuals for position ($\sim$60~mas) are still larger than expected based on simulations ($\sim$10~mas). We provide a likely explanation for the reason the small-scale residual errors are not corrected in our solution as would be expected.

\end{abstract}

\keywords{astrometry: general --- astrometry: individual}

\section{INTRODUCTION}

The prototype Panoramic Survey Telescope and Rapid Response System (Pan-STARRS1, hereafter PS1) is a wide-field imaging system, with a 1.8 m telescope and 7.7 deg$^2$ field of view, located on the summit of Haleakala in the Hawaiian island of Maui \citep[see ][]{kai10}. The 1.4 Gpixel camera consists of 60 CCDs with pixel size of 0.256 arcsec \citep{ona08, ton09}. It uses five filters (g$_{P1}$, r$_{P1}$, i$_{P1}$, z$_{P1}$, y$_{P1}$), similar to the ones used by the Sloan Digital Sky Survey \citep[SDSS ][]{york00}. The largest survey PS1 performs is the 3$\pi$ survey, covering the entire sky north of $-30\deg$ declination.

PS1 uses the 2MASS catalog \citep{skr06} as reference for astrometry, however this has been shown to introduce biases \citep{mil12, thol13} and large zonal errors (large-scale systematic errors on the sky).  When constraining the motion of asteroids, \citet{mil12} found residuals in both right ascension and declination that were both positive with 50-100 mas offsets.  \citet{thol13} suggested these large biases stemmed from uncorrected proper motions $-$ a direct result of using the 2MASS catalog as reference.  Later, \citet{far15} used proper motions from selected stars in the PPMXL catalog \citep{roes10} to make corrections to the PS1 solution (and other catalogs), successfully removing the large-scale zonal residuals from the  asteroid astrometric solutions.

Classical astrometry methods require the use of stellar-based reference catalogs (such as the 2MASS catalog in this example) making high-accuracy, bias-free absolute astrometry difficult. Zonal errors are very hard to remove from these stellar-based reference catalog.  Instead, post-processing `corrections' are made to compensate for obvious biases.  

A much cleaner concept of absolute astrometry is the global solution (GAS), better known as the block adjustment method (BA). Instead of using stars from a reference catalogs to tie each observation (so-called reference stars), this method uses an absolute reference frame, such as the International Celestial Reference Frame \citep[ICRF, see][]{icrf2}.  The ICRF2 is a zero proper-motion, zero parallax, higher-precision absolute reference frame from which to derive all 5 astrometric parameters ($\alpha$, $\delta$, $\mu_{\alpha}$, $\mu_{\delta}$, $\pi$) for each observed source in a given survey.  In essence, the 5 astrometric parameters for every observed source are `adjusted' simultaneously in a large least squares solution.  The advantage of this block adjustment method is that the observations are tied together into a more-or-less rigid block. In this case relatively few reference objects can be used to align the block of observations to a reference system and calculate absolute positions.  This approach is clean and rigorous but can be a much more complicated solution computationally (even prohibitively so for extremely large surveys).   In addition, the inputted data have very specific requirements.  For instance, a thorough cleaning of the data has to be done before setting up the block adjustment equations.  Introducing bad data to the solution can result in a null-result; the bad data being difficult to track after the fact. 

The first rigorous implementation of block adjustment was developed by \citet{eich60}.  The technique was developed further by several other authors \citep{googe70, vegt72, vegt74, vegt91}.  These methods use a first order expansion to calculate small shifts from an initial assumed solution. A different method that calculates sky positions directly was proposed by \citet{stock81}, but it does not provide proper motion and parallax.  Simulations or solutions for a limited amount of data have been performed in the past \citep{stock81, zac92, yu04} and the results were promising.  However, this work is the first attempt to use the block adjustment technique to derive an astrometric solution for a large catalog. 

For convenience we will use the terms `frame' and `plate' interchangeably for the full array (mosaic) of PS1 CCDs.  We use the term `observation' to describe the individual data taken for one star on one plate. For example we can say that one frame has 200 stars on it and that one star has 40 observations (appears on 40 frames).  Additionally, we will call our reference catalog the `quasars catalog' even if a significant fraction of the reference objects are actually Active Galactic Nuclei (AGN) and not technically `quasars'.  Because we are using astrometric positions obtained by Very Long Baseline Interferometry measured directly on the ICRF2, the quasar positions have accuracies of less than one mas; for purposes of our analysis, we thus assume the reference position error for these objects to be negligible. We also assume they have no proper motion or parallax since their distances are so large compared with the stars we are most interested in. Finally, `grid stars' are the stars to which the block adjustment is being applied.

In Section 2 we describe the data filtering, the grid and the quasar reference catalogs.  In Section 3 we present details of our BA algorithm and the results of simulations. In section 4 we present the results obtained with the PS1 data and validations. Finally in Section 5 we summarize the analysis and results.

\section{DATA, CATALOGS, AND REQUIREMENTS}

The block adjustment method requires that the input data meet certain density, uniformity and stability requirements.  We provide a brief overview of these requirements in the subsequent paragraphs, but provide many more details on the grid star and quasar catalogs used for this analysis in the following subsections. 

For grid stars, the first requirement is that the density of the stars be high enough on each plate that the field is solvable for the plate model chosen.  The grid catalog must also be as uniform as possible over the sky. The actual observations of the grid stars have similar requirements, both in number and uniformity and both in space and time. These requirements can be quantified by the number of observations per star (overlap factor) and by the time span for each star. If the overlap factor or the time span are too small, the plates will not be rigid enough for a solution and the system will be ill-conditioned. 

Globally, the catalog of quasars has similar requirements.  First, the link between the VLBI positions and the ICRF2 must be much better than the actual measurement error. This is straightforward if the quasar is an ICRF2 source, but for non-ICRF2 sources the link can be more challenging.  Second, a sufficiently high density of quasars in the observed $3\pi$ of the sky must be available to enable astrometric FoV calibration in the least-squares adjustment.  Third, the catalog must include a rather uniform distribution of reference sources on the sky to prevent local correlated error build$-$up.  Finally, optical counterparts should be in the magnitude range suitable for \PS. The ICRF2 catalog for example, is not suitable by itself, because it does not meet all of these conditions (e.g. it is not dense enough).  On the other hand, if the grid star density is high, then the frames are tied together in a relatively rigid system. This provides some latitude with quasar density, since the quasar density can then be much smaller compared to the grid stars. However the density of the grid stars ultimately limits the rigidity of the system and the plate model.  Therefore, having a relatively high number of quasars distributed uniformly is crucial for a global solution.

\begin{figure}
\epsscale{1.0}
\plottwo{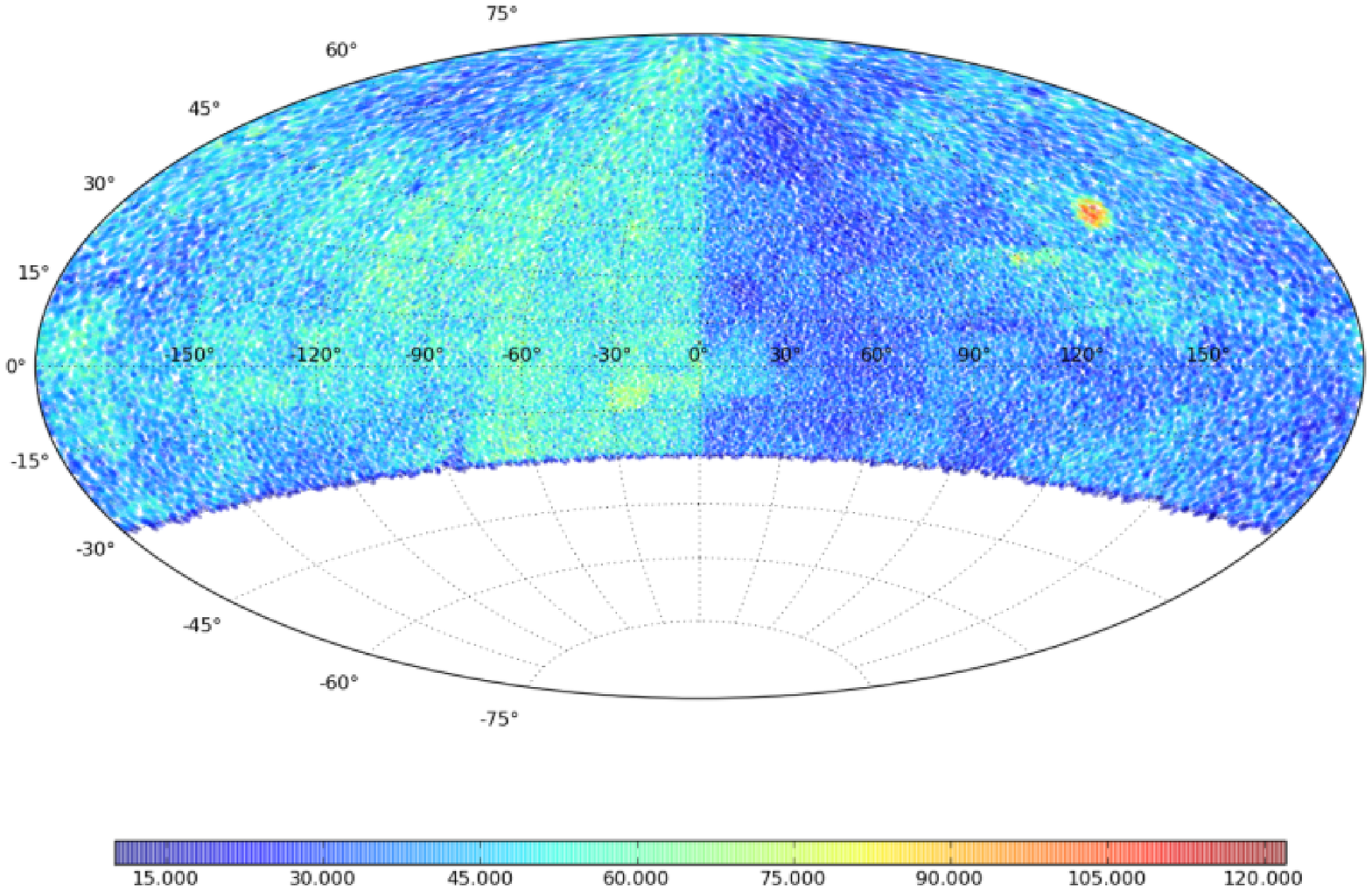}{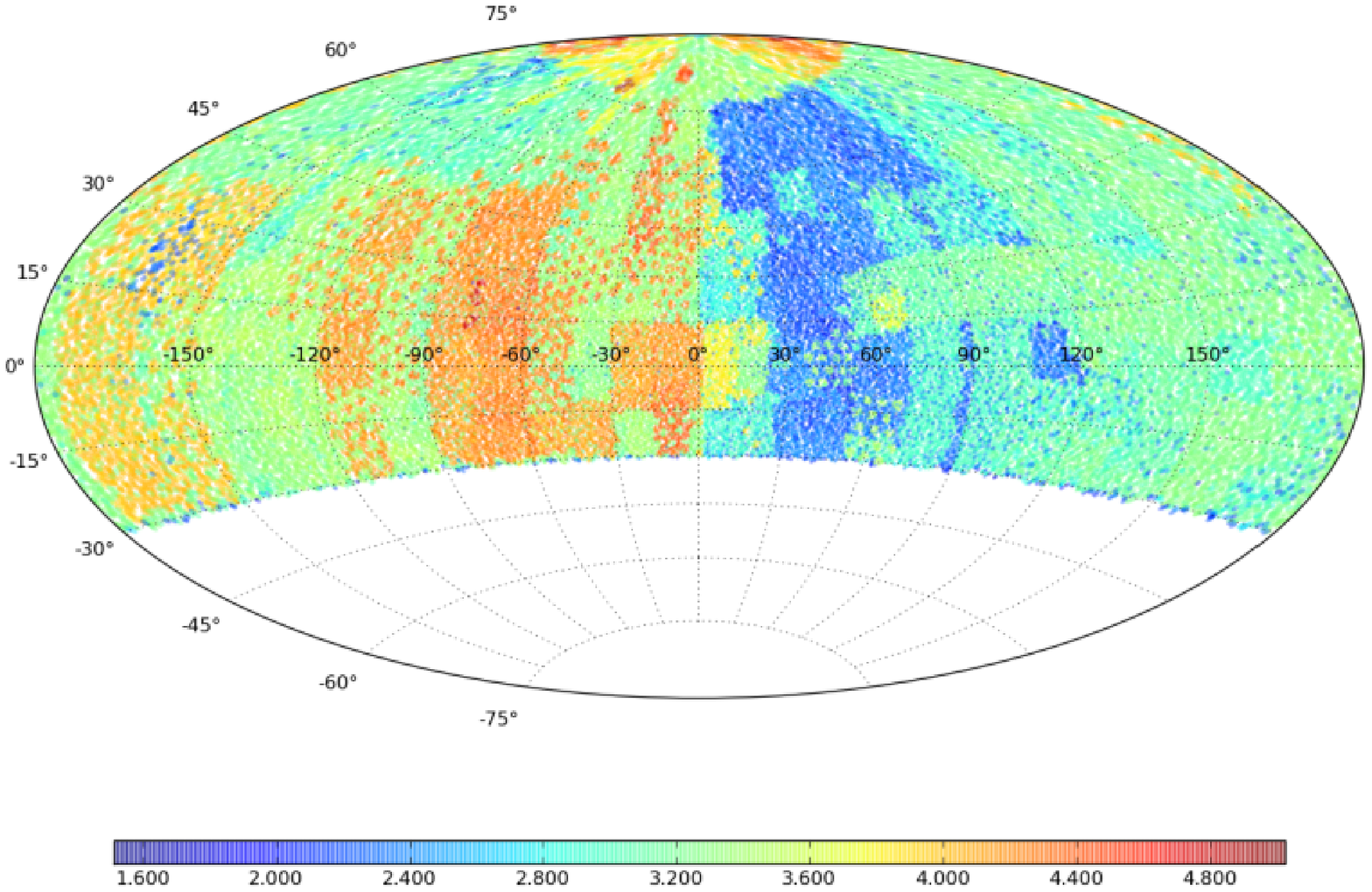}
\caption{
Observational overlap (left) and observational time span coverage (right) for the grid star observations.  Units are total number of observations (left) and total time span in years (right).
} \label{overtime}
\end{figure}

The PS1 data for the grid star and quasar catalogs described in the following subsections were obtained from the PV2 internal release with preliminary calibration statically tied to 2MASS. We used a cone search of one arcsecond from the catalog position. The data includes $\sim$30 million individual observations covering 3$\pi$ of the sky over 5 years with an average overlap factor of 38. We also obtained metadata which includes pointing, timing and filter information. The overlap and time span distribution on the 3$\pi$ sky is shown in Figure~\ref{overtime}. The data includes both `chip' coordinates (on each individual CCD) and `mosaic' coordinates (coordinates calculated with respect to the focal plane, or ``frame''; see section 1 for definitions). It also includes sky coordinates for each observation from the preliminary 2MASS-tied calibration.

\subsection{The Grid Catalog} \label{gridcat}

The global solution is a memory-intensive calculation.  The matrix solver, and the number of unknowns that can be solved (i.e. the star parameters) are limited by the amount of computer memory available. We therefore created a grid catalog of 750,000 stars based on the UCAC4 catalog \citep{ucac4}, which does not go as deep as Pan-STARRS but has enough stars to meet the density and uniformity requirements described above. 

\begin{figure}
\epsscale{0.7}
\plottwo{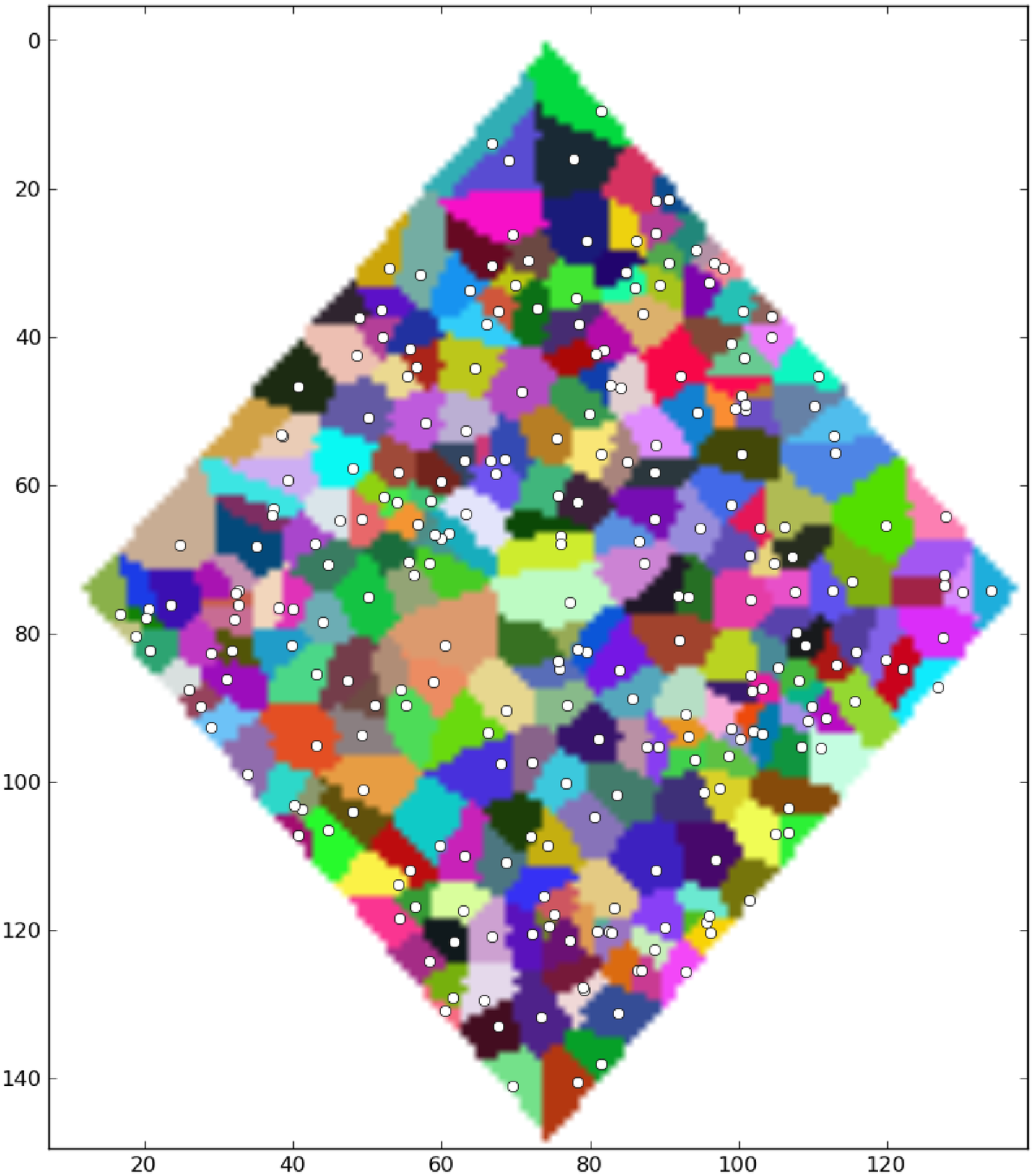}{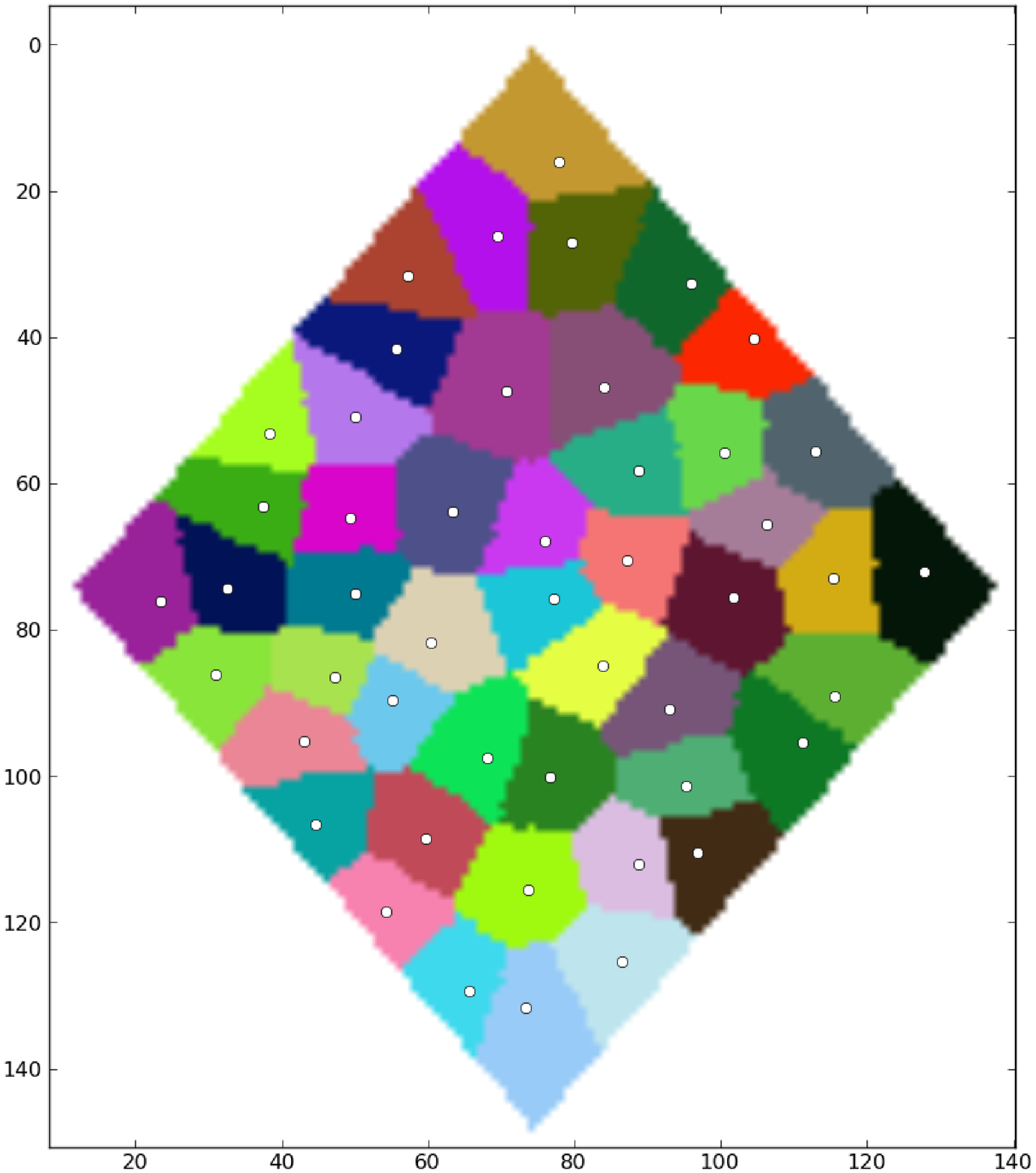}
\caption{
Example of Voronoi tesselation algorithm used to generate the grid catalog. The sky was broken into $\sim$~0.8 square degrees Healpix cells, like the one shown here. We start with the UCAC4 stars on the left and apply the Voronoi tesellation. Then stars with the smallest Voronoi area are removed iteratively until a fixed number of stars remains in each cell, as shown in the right image.
} \label{voronoi}
\end{figure}

We used HEALPix \citep{gor05} to partition the sky in small parts of equal area ($\sim$~0.8 square degrees for nside=64). We also made a magnitude cut to the UCAC4 catalog, requiring all stars be fainter than 15.5 magnitudes.  We then used an algorithm based on Voronoi tesselation to assure uniform sampling of stars in each small partition.  In this process, a Voronoi cell is constructed around each star by straight lines equidistant between each  pair of neighboring stars and perpendicular to the line that connects them.  The area of this cell is larger if the star is far from its neighbors. To obtain the optimal distribution, the stars within the smallest Voronoi areas are removed iteratively until a fixed number of stars in each of the small HEALPix regions are obtained. One example is shown in Figure~\ref{voronoi}. On the left we show the original population of UCAC4 stars on one of the HEALPix regions with the Voronoi partitioning while the right image shows the final, uniform selection of our grid stars. The Voronoi partitions are now almost equal in size, which indicates that the star distribution is uniform. 

The resultant grid star catalog has an average density of 24 stars per square degree for the region observed by PS1.
Finally we note that once we solve for the grid stars, they can be used as reference stars to obtain a relative solution for the rest of the stars in the PS1 data. 

\subsection{The Quasars Catalog} \label{icrfcat}

To develop our quasar catalog, we used the OCARS ({\it Optical Characteristics of Astrometric Radio Sources}) catalog \citep{ocars, mal,tit}.  The OCARS catalog is a carefully maintained compilation of extragalactic radio sources with accurate VLBI positions, cross-matched in the optical and NIR passbands. At the time of our grid catalog preparation, OCARS included 9027 separate sources, now it counts 9392 separate entries\footnote{The source file is at  \url{www.gao.spb.ru/english/as/ac_vlbi/ocars.txt}.}. Most of the new additions are coming from the steadily growing Radio Fundamental Catalog (RFC)\footnote{\url{http://astrogeo.org/rfc/}}.  In addition to accurate VLBI positions on ICRF2, OCARS includes redshifts $z$, when available, morphological classification, photometric data on optical counterparts in an annex, and a useful cross-identification table. With a circular search area of $1\arcsec$, we find 5034 RFC sources detected at least 10 times by PS1. This implies a 74\% rate of optical identification for VLBI$-$observed radio sources. The cross$-$matched objects represent the preliminary basis of our reference sample.

\subsubsection{RORFO}

\label{rorfo.sec}
Optical counterparts of radio sources with accurate VLBI positions should be carefully vetted before they obtain the status of Radio-Optical Reference Frame Objects (RORFO). This is especially important for our application, because even a small number of strongly perturbed or mismatching sources can bring about local areas of large position error. Constructing a global (or nearly global in our case) astrometric grid from small$-$field differential observations is inevitably fraught with poorly conditioned large$-$scale correlated errors, which can be viewed as a ``red" spectrum of absolute error realization in terms of orthogonal spherical harmonics \citep{mak05}. The ways of stemming this dangerous build$-$up of large$-$scale distortions include using a wide basic angle between two fields of view \citep{mak}, as in the Hipparcos and Gaia missions, a rather wide ``field of regard" with internal regularization, as in the {\it Space Interferometry Mission} \citep{unw}, or, as we do it here, an absolute reference grid of extragalactic sources \citep{mado}. Having resolved the problem of large-scale error, we are confronted now by the danger of medium$-$ and small-scale perturbations of the solution, which may get out of control on the scales corresponding to the typical separation between the reference objects. If the observed optical position of one quasars is far from the assumed radio position (100~mas or more), a large local perturbation occurs and propagates into the global solution, pulling the results for all other stars in the surrounding area.

The density of the reference grid is limited by the number of available cross-matched VLBI sources, which can not be drastically improved in the near future. The quality and the reliability of the sample becomes most important. The first step of reference sample cleaning was to visually review a large number of digital images available for the brighter part of the sample and reject any sources that do not look compact, symmetric, and point$-$like on the sky. Extended, double and perturbed galaxies with dust structures are especially common among the brighter VLBI counterparts. For example, 23 of the RFC sources have \PS counterparts with $i_{P1}$ magnitudes brighter than 13; 21 of these objects were rejected as RORFO. In the next magnitude interval, $i_{P1}\in [13,14]$, 8 out of 16 objects were filtered out. The rate of obviously unsuitable sources further drops to 35\% for $i_{P1}\in [14,15]$, 20\% for $i_{P1}\in [15,16]$, and 11\% for $i_{P1}\in [16,17]$. The main reason for this tendency is the fact that the optically brighter host galaxies of AGNs are nearby, and therefore better resolved in the available images of a limited seeing.

\subsubsection{Radio-optical offsets} \label{offsets}

The remaining sources are considered candidate reference objects, but the distribution of the PS1$-$VLBI position offsets reveal the presence of a large number of possible mismatches and problem cases. At the preparatory stage of our astrometric solution, we made use of preliminary positions for our grid objects (see the begining of this section) computed by the pipeline at IfA. The original intention was to filter the most obvious PS1$-$VLBI mismatches and to investigate the possible reasons. The procedure turned out to be more complicated than what had been expected, and additional data processing methods had to be engaged. The preliminary positions suffered from considerable large-scale sky-correlated errors, which can be best represented by a set of nearly-orthogonal, low-order vector spherical harmonics \citep{mamu,kli}. These large zonal errors are probably related to the uncorrected proper motions from the 2MASS reference catalog, as explained in the introduction.

A spherical harmonic fit to 7th degree on the 4979 VLBI sources from the RFC remaining after the discarded objects elimination (Table \ref{rej.tab}), revealed a statistically significant vector field dominated by a few low-order dipole harmonics. The magnitude of the sky-correlated error reached $\sim 70$ mas in some parts of the PS1 sky. We remove the sky-correlated perturbation using 96 vector spherical harmonic functions and analyze the post-fit residuals. The median magnitude of the fitted error is 55~mas, which is an estimate for the PS1 errors if the large biases described in the introduction are removed.

\begin{figure}[htbp]
\epsscale{1.0}
\plottwo{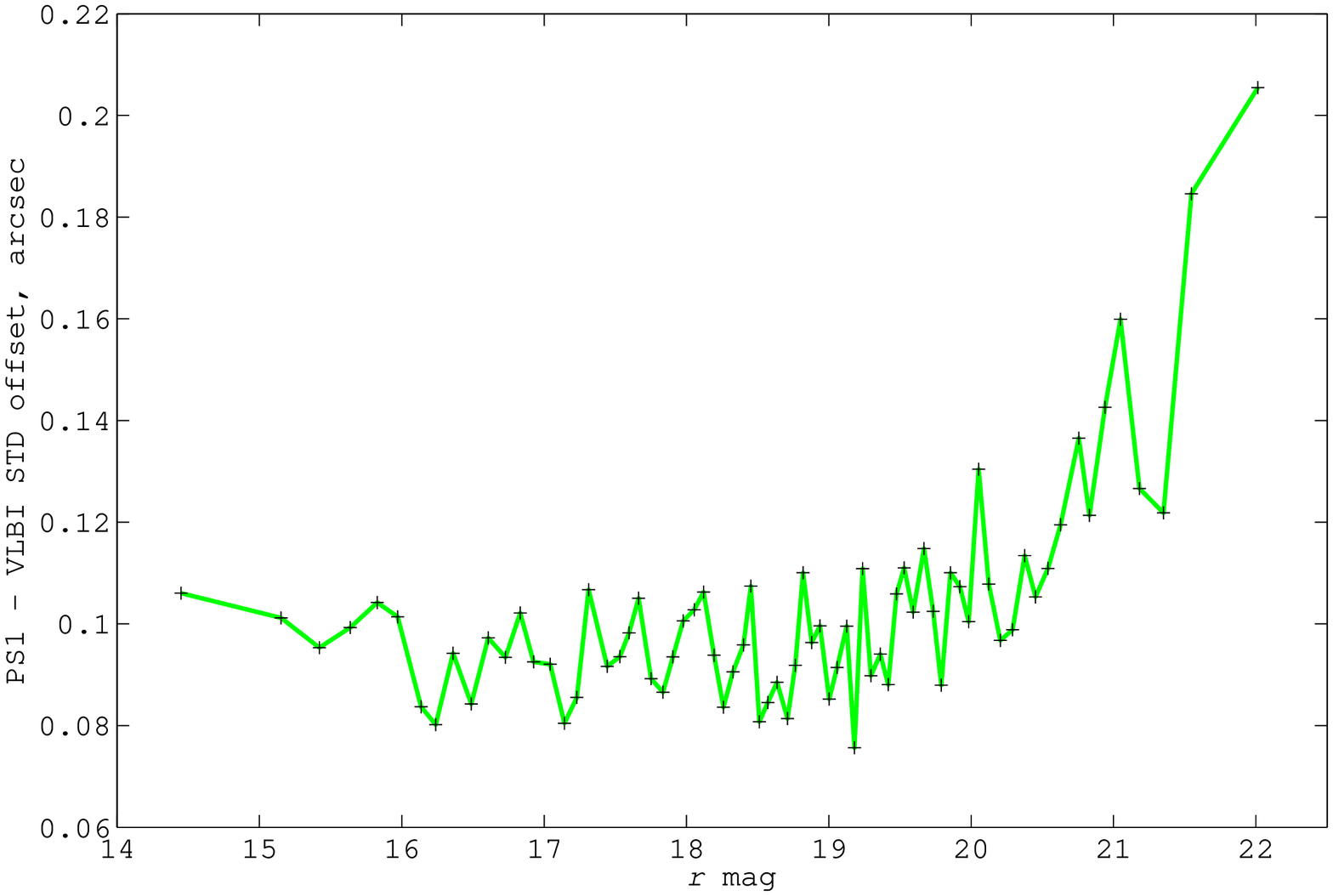}{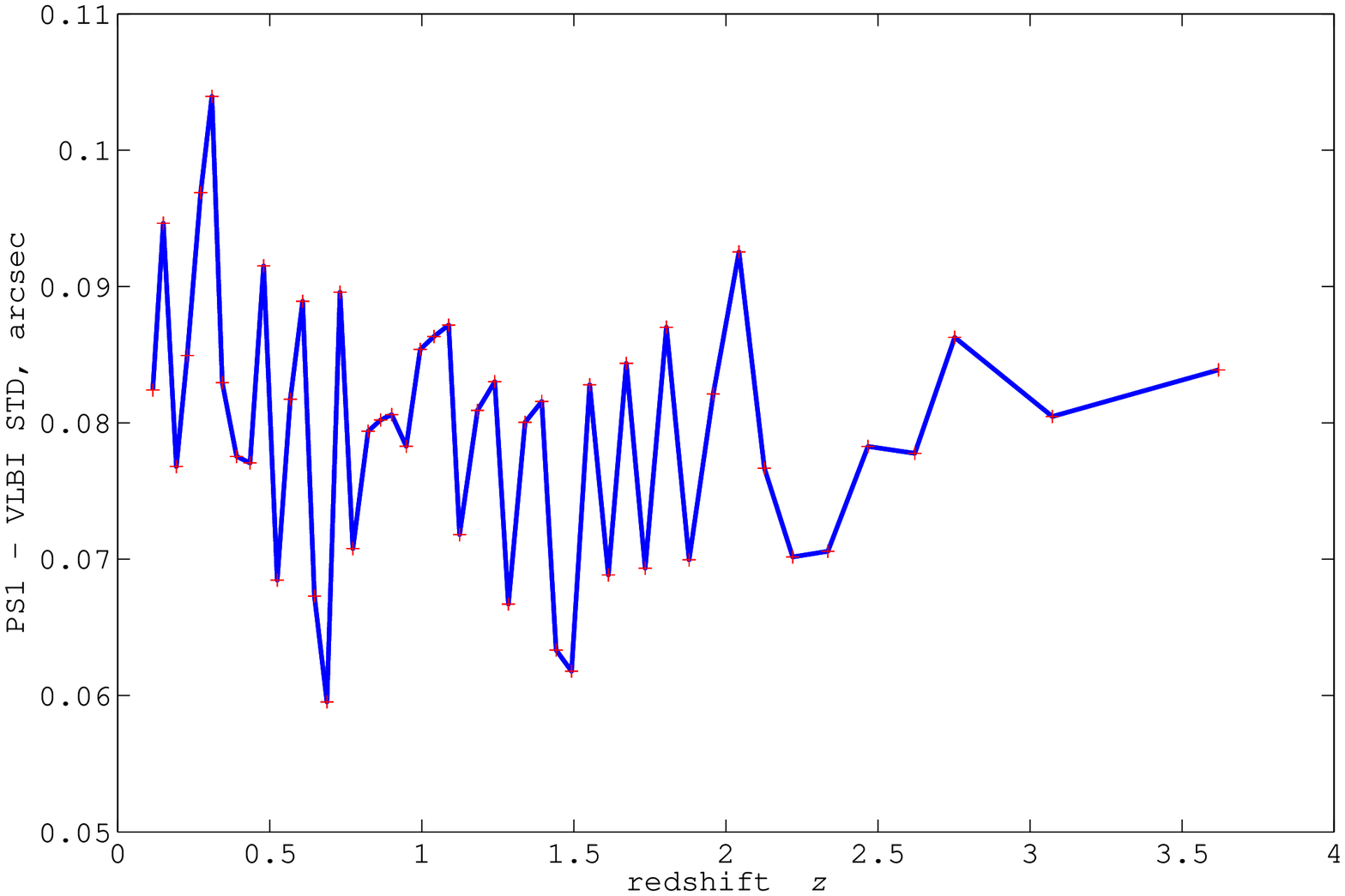}
\caption{Standard deviation of PS1 $-$ OCARS position differences as a function of r$_{P1}$ magnitude (left)
and as a function of $z$ redshift (right). \label{std.fig}}
\end{figure}

Fig. \ref{std.fig}, left shows the standard deviation of the offsets binned by the observed r$_{P1}$ magnitude for the 4979 extragalactic sources. The residual RMS scatter is flat for magnitudes between 16 and 19 at approximately 90 mas, which we consider to be the initial error of positions before our global solution. The standard deviation begins to turn up for objects fainter than $r_{P1}=19$ and rapidly rises at $r_{P1}>20$ mag. Photon-limited astrometric precision is characterized by an exponential rise of random error with magnitude; here the Poisson shot noise becomes visible only at the faint end of the range. Therefore, for the majority of the reference objects, the sources of error are other than photon statistics. A small rise of offsets seems to be present also for the brightest objects, $r_{P1}<16$ mag. This may be interpreted as a higher rate of resolved, extended objects at brighter magnitudes. Generally, the AGNs in nearby galaxies can be expected to be brighter. We should see a similar build-up of random offset for objects with smaller redshift. On the other hand, as was speculated in \citep{maq} based on the empirical ``fundamental plane" relations found by \citep{ham}, the QSO observed at higher $z$ should have higher nucleus to host brightness ratios, and thus, statistically smaller radio-optical position offsets. Fig. \ref{std.fig}, right confirms this, as the nearer VLBI sources at $z<0.6$ have larger radio-optical offsets than the objects at higher redshifts. Ideally, we would like to use only objects with redshifts $z>0.6$, but this depletes the number of available reference objects below the critical value. Note a curious difference between the average level of standard deviations in Fig. \ref{std.fig}, on the right it is smaller than 80~mas. Only some of the OCARS sources have their $z$ determined and listed, and those tend to be less offset with respect to the VLBI positions.

\subsubsection{The origin of large radio-optical offsets} \label{origin}

\citet{zz} empirically detected an increased scatter in the radio-optical position differences for a sample of brighter QSOs above the expected random error level and suggested that {\it all} optical counterparts are astrometrically perturbed at the level of 10~mas. The astrometric accuracy of the PS1 data does not allow us to test this surmise, but we find a significant number of large offsets, extending up to 1 arcsec (limited by our search radius), even after the visual culling described in \S \ref{rorfo.sec}. Are these differences real, or just evidence of a strongly non-Gaussian distribution of random error in PS1?

\begin{figure}[htbp]
\epsscale{0.7}
\plotone{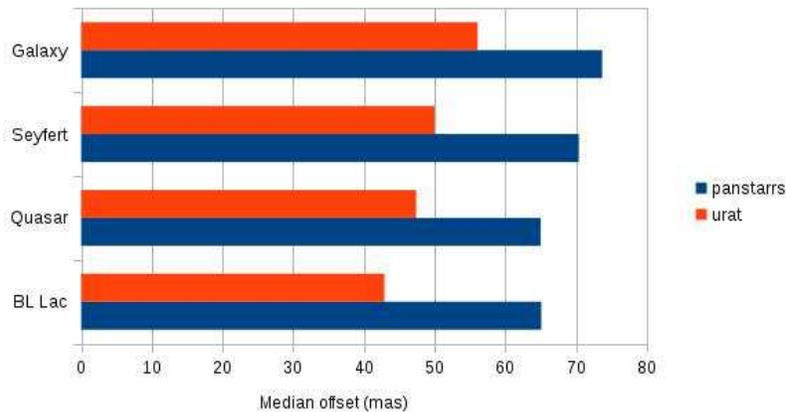}
\caption{Median position offsets PS1$-$VLBI and URAT1$-$VLBI for major morphological classes in OCARS, in mas. \label{hist.fig}}
\end{figure}

\begin{figure}[htbp]
\epsscale{1.0}
\plottwo{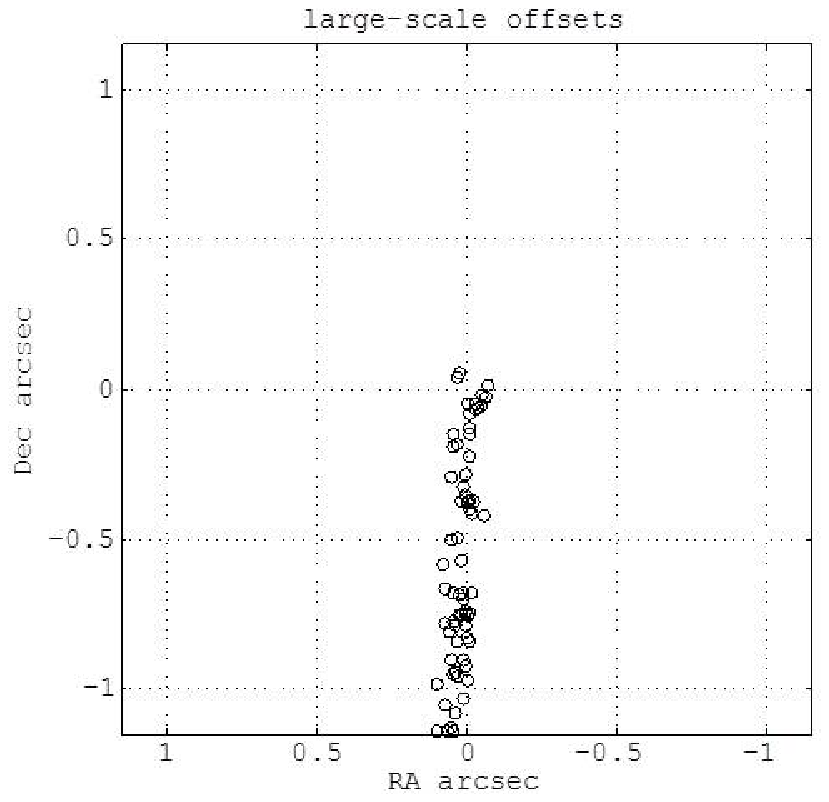}{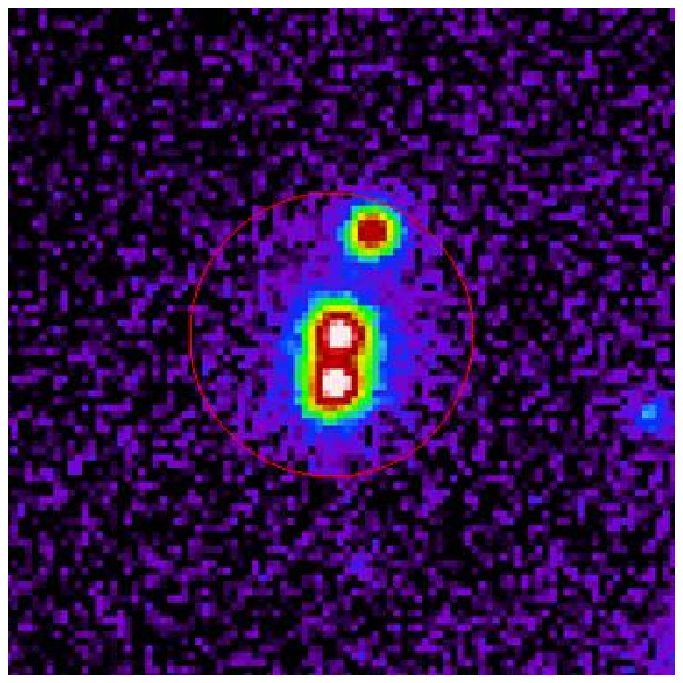}
\caption{The VLBI source VCS3 J2137+3455 is identified as a compact symmetric object (CSO) in radio.
Left: the cloud of astrometric detections in PS1. Right:  the Pan-STARRS composite image in the i$_{P1}$ band reveals two galaxies sseparated by 2 arcsec. The radio source corresponds to the northern component of this system. The red circle of 5 arcsec radius is centered on the VLBI position of the radio source.\label{double}}
\end{figure}

First, using the morphological classification of optical counterparts in OCARS, we can compare the typical scatter of offsets for different types of objects. Fig. \ref{hist.fig} represents the median magnitude of offsets for the major types: quasars, galaxies, Seyferts, and BL Lac-type objects based on preliminary PS1 data and the USNO Robotic Astrometric Telescope (URAT)-1 catalog \citep{urat}. Quasar and BL Lac counterparts provide more consistent optical positions than Seyferts and especially, galaxies. The reason for this segregation is fairly obvious: optically bright host galaxies are often asymmetric in  shape due to merger events, duplicity, or prominent dust structures. For example, the galaxy NGC 5675 has a compact radio-loud AGN \citep{pus}, which is an ICRF source J143239.8+361807, but the available HST WFPC images reveal a large inclined galaxy with a powerful dust lane. Since the dust structure is tilted with respect to the line of sight, the obscuration on the two sides of the image is asymmetric. As a result, the PS1 detections are shifted by almost 400 mas at position angle $52\degr$, which is very close to the axis of the dust lane. This galaxy is well resolved because it is nearby, whereas a more distant analog would look small and fairly compact in the optical images, still producing a measurable astrometric displacement. Another interesting example is the high-quality VLBI source VCS3 J2137+3455 \citep{pet}, whose PS1 detections are strongly perturbed and lined up in almost exactly North-South direction, extending to 1 arcsec off the VLBI position. The digital \PS image (Fig. \ref{double}) reveals a double galaxy with two components of nearly equal brightness separated by $\sim 2\arcsec$. The double galaxy is seen as the two brighter sources at the center of the image (the third fainter source is an unrelated foreground star). The radio-loud source corresponds to the northern component. At a much smaller scale on the radio images, the source was identified as a compact symmetric object \citep[CSO,][]{sok}. 

\begin{figure}[htbp]
\epsscale{1.15}
\plottwo{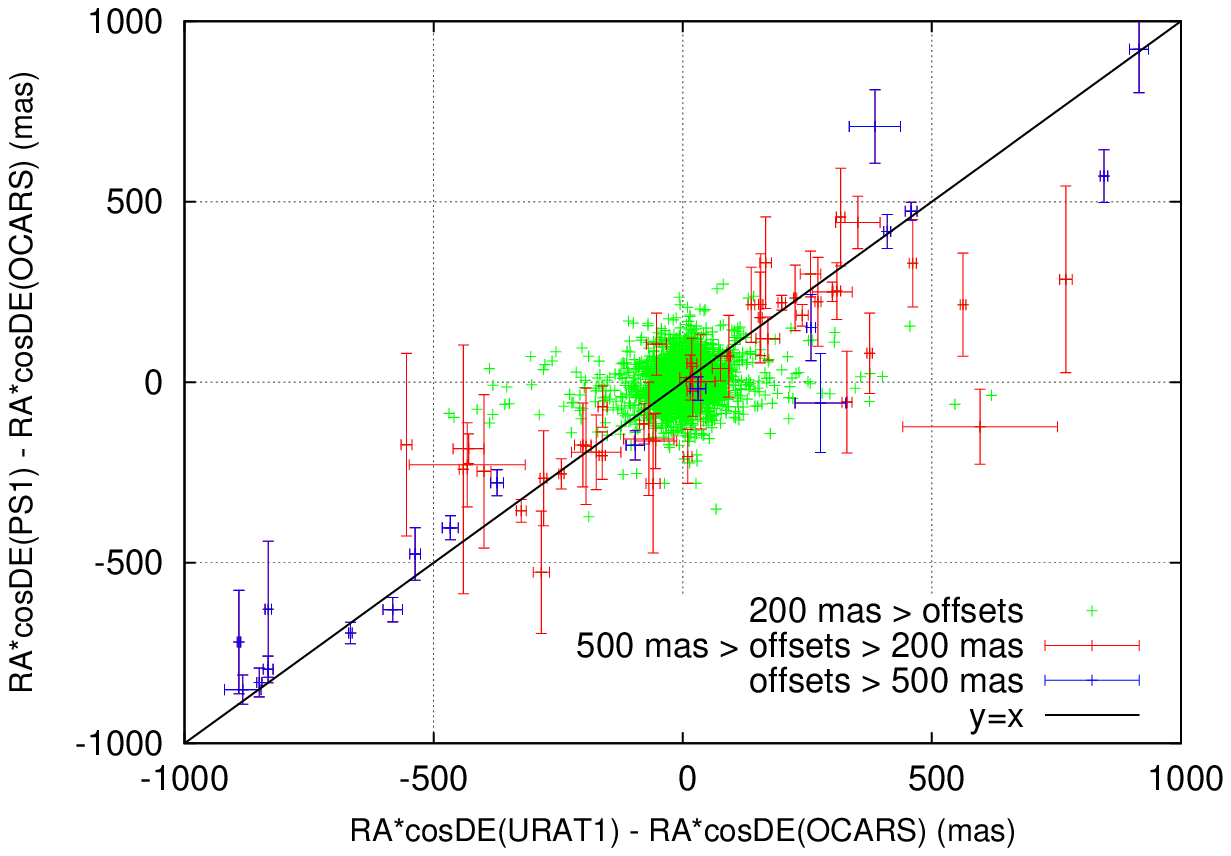}{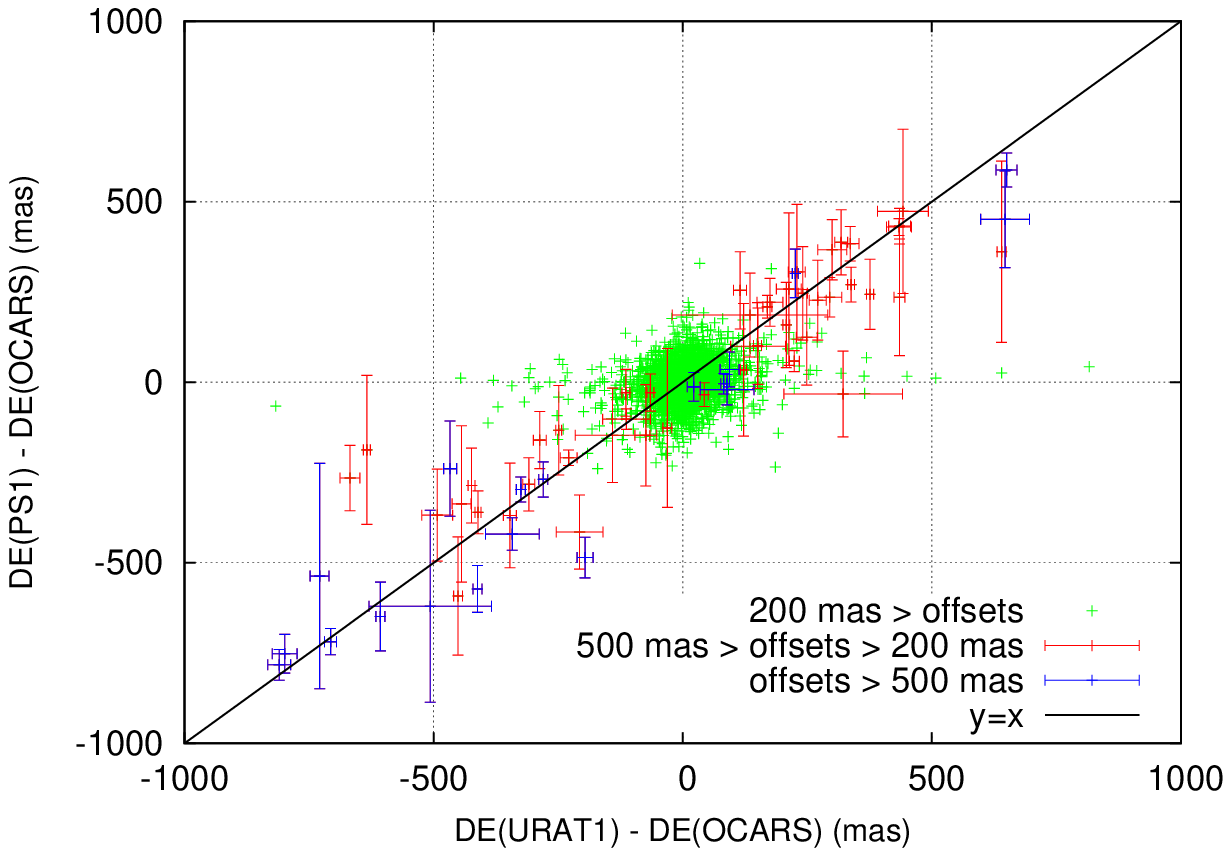}
\caption{Radio-optical position offsets PS1$-$VLBI versus URAT1$-$VLBI in RA ($\Delta\alpha\;\cos \delta$, left)
and Dec ($\delta$, right). \label{ps1vsurat.fig}}
\end{figure}

Second, we can verify the large offsets of preliminary PS1 positions using the URAT catalog for brighter northern objects, which is a completely independent astrometric catalog of comparable or superior accuracy. Fig. \ref{ps1vsurat.fig} shows a clear correlation between the PS1$-$VLBI and URAT$-$VLBI offsets for common objects with PS1$-$VLBI offsets greater than 200 mas. The majority of such large offsets is real and physical. In some cases, we could trace the cause of the large offset using available high-resolution images, \PS images, multi-band epoch photometry, and the distribution of astrometric detections. The most frequent causes are 1) asymmetries in the resolved host galaxies; 2) double sources and image blending, including double galaxies, mergers and optical stellar companions; 3) microlensed systems. The object PKS B2114+022 is an example of the latter category, being a compact bona fide VLBI source, for which the available HST NICMOS images show two compact galaxies beside the radio position, possibly related to the microlensing but not to the AGN. The optical counterpart of the radio core is not visible at all in this case.

\subsubsection{Final Quasar Catalog}

\begin{figure}[htbp]
\epsscale{0.6}
\plotone{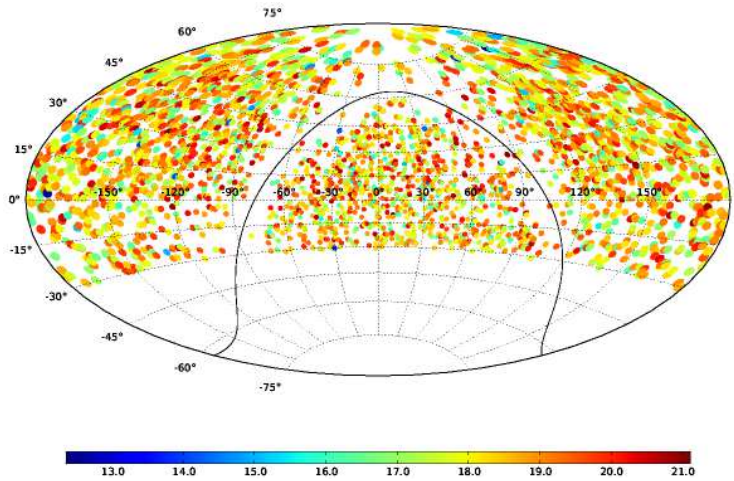}
\caption{The Quasars Catalog positions with average g$_{P1}$ magnitudes.}
\label{ocars}
\end{figure}

Based on results in \S \ref{origin}, we decided to keep only sources classified as `BL Lac' and `Quasars'. This removes 2670 objects classified as `Radio sources' mostly located in the galactic plane and 1327 `Galaxies'. As shown above in \S \ref{offsets}, many of these sources have large offets from the radio positions due to asymmetric shapes. We remove the sources with large offsets based on URAT (\S \ref{origin}). A total of 57 sources were removed based on visual inspection and URAT comparison. The removal of all these sources produced big gaps along the Galactic plane. In order to mitigate this problem we re-inserted sources with redshift larger than 0.1 in OCARS, the reasoning being that galaxies at large distances have a smaller angular size and therefore the offests between the radio and optical are likely to be smaller. Their positional accuracy is probably not as good as the other sources but they provide important constraints close to the Galactic plane, without which a global solution would not be possible. The final quasars catalog is shown in Figure~\ref{ocars} and contains 3076 sources.

\subsection{Data filtering} \label{filters}

As discussed at the beginning of this Section, the GAS method has certain requirements for well-behaved data in terms of density, uniformity and stability. Therefore a rigorous filtering is necessary to remove data which could yield an ill-conditioned system. If the solution fails after the input data are choosen it is very hard to trace these `bad data' after the solution has failed. Therefore the filtering has to be as rigorous as possible before performing the global solution.    Several types of filtering were performed, including removal of binaries, positional outliers and objects with insufficient observations.  We describe the filtering procedures in detail below. 

\begin{figure}
\epsscale{0.7}
\plotone{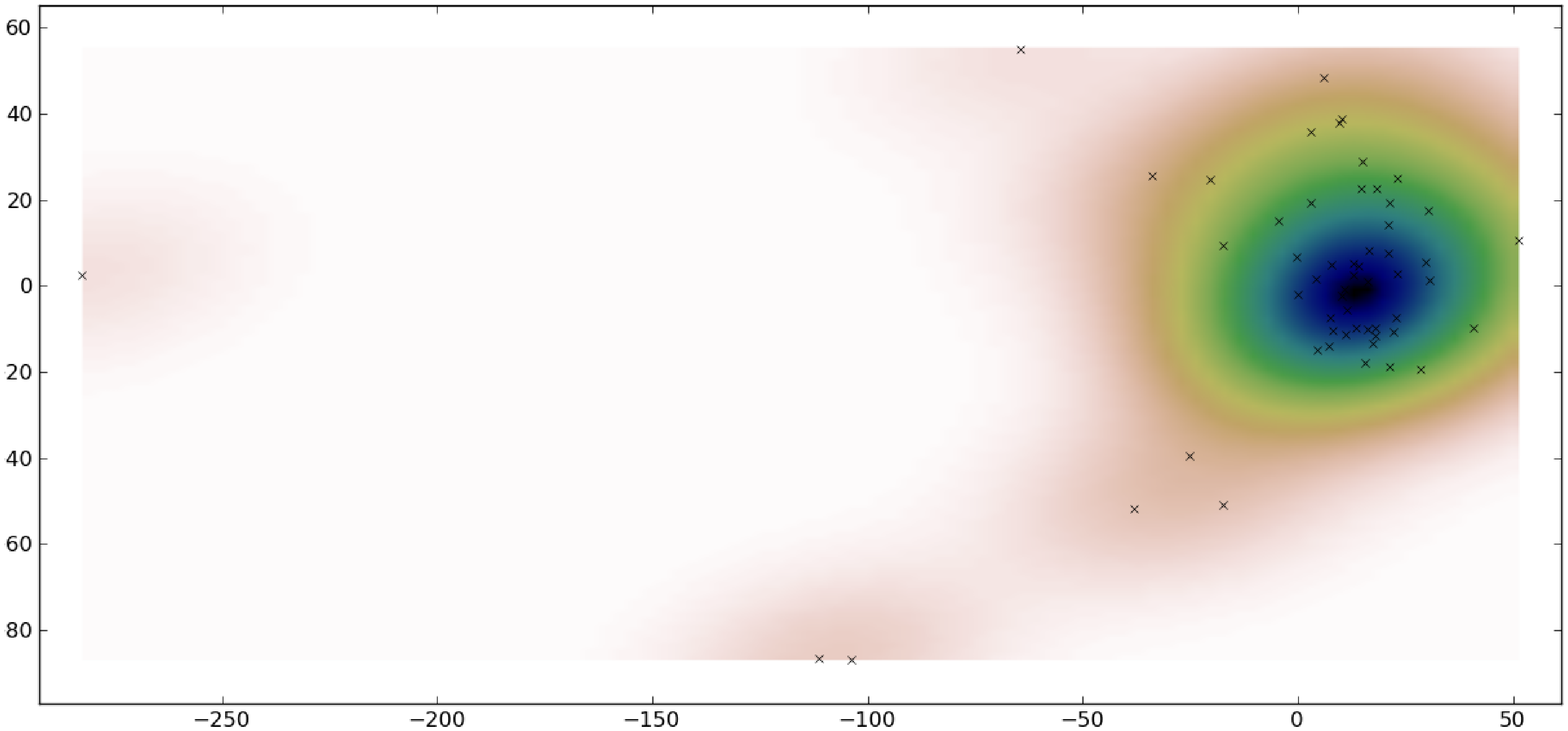}
\caption{
Example of outliers in the PS1 pipeline solution. The image shows the individual sky positions for one star. Seven of them are outliers and are removed in the filtering process The units are offsets (in mas) from the mean position.} 
\label{outliers}
\end{figure}

First we removed duplicate detections produced by the 1 arcsec cone search (see the beginning of this Section). Some of these duplicates are probably binaries and we attempted to remove binaries to avoid the astrometric complications associated with these sources for the initial solution. Next we removed positional outliers, which are those sources that do not match the sky positions provided with the PS1 data (based on 2MASS).  In other words, most observations are located in a cluster with more or less random positional errors but a few observations are `positional outliers'.  These outliers are likely caused by source confusion or some unknown error in the PS1 reduction pipeline.  To remove positional outliers we used an algorithm based on a Gaussian kernel density with automatic bandwidth determination to remove observations outside a two-dimensional probability density of 2~$\times$~10$^{-5}$. We note that such filtering has the potential to affect proper motion results, but this was not the case here according to our tests (see the next section). One example of the outlier filtering is shown in Figure~\ref{outliers}. 

In order to obtain a system of equations that are well-conditioned we require that each frame contain at least 20 stars, and that each star be observed at least 10 times. It is also important that the stars are distributed uniformly on the plate, so in addition we require that each quadrant of each frame contains at least 5 stars. It is still possible that the stars are concentrated around the center of the frame but this probability is low given the method we used to construct a uniform grid catalog (see \S \ref{gridcat}). We chose this simple requirement for speed; otherwise a more complicated algorithm would have been required. In order to constrain the proper motion we also impose a time span limit of at least 1.5 years of observation for each star. 

The filtering process has to be done iteratively. For example if a star is removed because it only has 9 observations or because the time span is only one year, then we have to go back and count the stars again on all the frames where the star was observed. Similarly, if one frame is removed because it has only 15 stars then for each of these stars we have to count again the number of observations and calculate the time span. The imposed conditions for this filtering process are adjusted based on the quality of the data, such as the overlap factor and the time span (see Figure~\ref{overtime}). The conditions must  be chosen so that the iterative process converges and that no gaps in the data are  introduced.

\section{ALGORITHM} \label{ba}

The equations that relate the sky coordinates of a star and its position on the detector (x, y) are not linear. The Eichhorn approach (see the Introduction and references therein) is to linearize the equations in a Taylor expansion, keeping only first order terms.  In so doing,  {\bf small shifts from the assumed values of the unknowns} are calculated. We assume that the zero values are known to a reasonable accuracy (better than 1 arcsec) so that only one step is required to calculate the shifts. In this approach, the equations coefficients are the Jacobian of the detector positions with respect to the unknowns. We present the exact details of this approach in Appendix A.

Once the coefficients are calculated for each star and each frame, they are assembled into a so-called `design matrix' which has billions of nonzero elements. Solving such a matrix is challenging for even the best available matrix solvers running on extremely powerful computers. It is typically useful to eliminate one set of parameters to facilitate the solution. There are two different ways to accomplish this:  via QR Elimination and/or via Block Elimination. In our code we use the QR factorization to eliminate the plate parameters plate by plate \citep{mak05}.  The Block Elimination method removes the star parameters after the design matrix has been constructed, using matrix operations and exploiting the shape of the design matrix \citep{vegt72}.  In Appendices B and C we present the two elimination methods in detail. 

After removing the plate parameters the matrix is normalized and then solved with the MKL (Math Kernel Library) PARDISO (Parallel Direct Sparse Solver Interface) solver \citep{sch04}. This is a very robust and fast multiprocessing parallel direct sparse solver. It supports out-of-core (OOC) option to store the matrix on disk instead of memory. This allows larger matrices to be solved, though at the expense of processing speed. All the code was run on a Blade server with two Intel Xeon X7560 8-core processors having Hyper-Threading and 512~GB of memory. The Blade allows 32 processes to run concurrently.

\subsection{Description of the code}

Our GAS code is Python-based and uses NumPy for high efficiency and speed. Most of the code modules are run in parallel, including the QR Elimination. The data are split beforehand using HEALPIX with an `nside' parameter of 8. This is useful for parallelism and to speed up simulations (see the next section).

The first part of the code is sequential: the data and metadata are read for each frame and converted to a format required by the BA equations. These are stored as memory maps which are later read by separate running processes to construct the design matrix. In this initial step we also create data structures for the grid stars and quasars and for the frames. Additionally we perform the filtering as described in \S \ref{filters}. As mentioned previously, some filters (for binaries, outliers, etc.) are performed in one step. The filters required for a well-conditioned system (removing stars with few observations and frames with few stars) are then run iteratively until all the `bad data' are removed. It usually takes only a few steps for all filters to converge and eliminate all the bad data. If the convergence is slow the adjustment parameters can be tightened, but care must be taken that no field gaps are introduced by removing too much data.

Now the data are ready for parallel BA. Each process reads its assigned data memory maps, calculates the equation coefficients frame by frame, performs the QR Elimination and constructs the design matrix. To save memory each process splits the design matrix into a number of slices which are later combined after the small matrices are normalized. Finally the normalized matrix is solved with PARDISO and the solution (the adjustments to the star parameters) is converted to sky positions and proper motions.

\subsection{Simulations} \label{sim}

Our GAS code includes a simulator, which can create realistic simulated data. It can simulate star catalogs and metadata (such as pointing, time stamp, orientation on sky). Observations are simulated by projecting the stars on the focal plane, given the instrument properties (telescope, camera, observing strategy). It can simulate measurement errors and field distortions on each frame. The simulator is very useful for debugging and testing purposes because actual data can be compared with simulated inputs. It also provides useful information about filtering `bad data', and provides a good estimate for the random and systematic errors we can expect.

In order to make the simulations as accurate as possible we based the simulated data on the actual PS1 metadata, quasars and grid catalogs (UCAC4 positions and proper motions). This ensures that we include any systematic errors, such as those caused by a non-uniformity of the stars on the sky, or by the observation strategy (overlap factor or time span). In fact the only difference from the real data are the measurement errors, possible plate distortions and other unknown systematic effects in the real data. The stars in our grid catalog are relatively bright (UCAC4 magnitudes in the narrow range 15.5-16.0). Random errors with single measurement of 50~mas or better for the over-sampled PSF (1.1 arcsec FWHM) of PS1 are expected for the PS1 data \citep{mil12}. Therefore we added random errors with a standard deviation of 50~mas to all the simulated measurements as well.

Given such large expected measurements errors compared to the average parallax (few mas), we can solve for positions and proper motions ($\alpha$, $\delta$, $\mu_{\alpha}$, $\mu_{\delta}$) but not for parallaxes. The attitude parameters for the observations and the calibration parameters (such as scale), are treated as nuisance parameters and eliminated  (see the beginning of this section and the Appendix). The simulation results indicate that the errors should be $\sim$10~mas for position and $\sim$9~mas yr$^{-1}$ for proper motion in each direction. We discuss these results in more detail in Section 4 below.

\section{RESULTS}

\subsection{The PS1 solution}

The PS1 CCDs are not absolutely fixed with respect to the focal plane array (FPA) and their position can shift slightly. Therefore ideally we would use the chip coordinates for this analysis so that we do not introduce errors related to the chip positions, which in principle could be significant. Unfortunately this would require a very large number of grid stars (2.3 million for a minimum of 10 stars per chip), which in turn requires a huge increase in computing capacity (memory). We therefore used the mosaic  coordinates to set up the equations. We emphasize that using mosaic coordinates instead of CCD coordinates can introduce potentially large errors related to the chip movement and other systematic effects as we will see later. Our definition of a frame is the full FPA, and the average number of stars per frame is then $\sim$~190.

First we filter the data as explained in section \ref{filters}. We use a simple first order plate model with only 6 parameters (see Appendix A for a description of the model). We use the UCAC4 positions as starting positions for convenience (any starting positions which are close enough for our first order approximation can be used). This choice was preferred over using average PS1 starting positions because it provides better validation methods as we will see below. Our solution calculates shifts from these starting positions. To get sky positions at the epoch 2012.6 (median epoch for the PS1 data) the calculated shifts are added to the UCAC4 positions. Simultaneously we calculate absolute proper motions (the assumed initial values are in this case zero). The plate parameters are eliminated as explained at the beginning of Section~3.

Validating the solution and estimating errors for real data is more complicated than when using simulated data. It requires the inversion of the design matrix, which is much harder to achieve computationally than a least square solution. The MKL PARDISO solver does not have such capabilities. Other options can be used to validate our solution and these are explored next.

\subsection{Position Validation}

\subsubsection{Using Quasars}

Quasars provide a good method to estimate absolute errors in star catalogs \citep[e.g.][]{urat}. However the quasars are used in GAS function as hard constraints (we do not solve for them) and therefore these cannot be used to estimate absolute errors. However, one method of validation is to remove a small fraction of the quasars in the quasar catalog  (758, a quarter of the total) and incorporate these into the grid catalog $-$ thereby allowing us to solve for this quasar subset as though they were grid stars.  These so called `validation' quasars were chosen randomly over the sky except in the galactic plane where quasars are already inadequate. Using a smaller number of quasars in our solution (3/4 of the total) may overestimate the errors since the solution has fewer hard constraints, but this methodology provides a good upper limit. 

\begin{figure}
\epsscale{0.9}
\plottwo{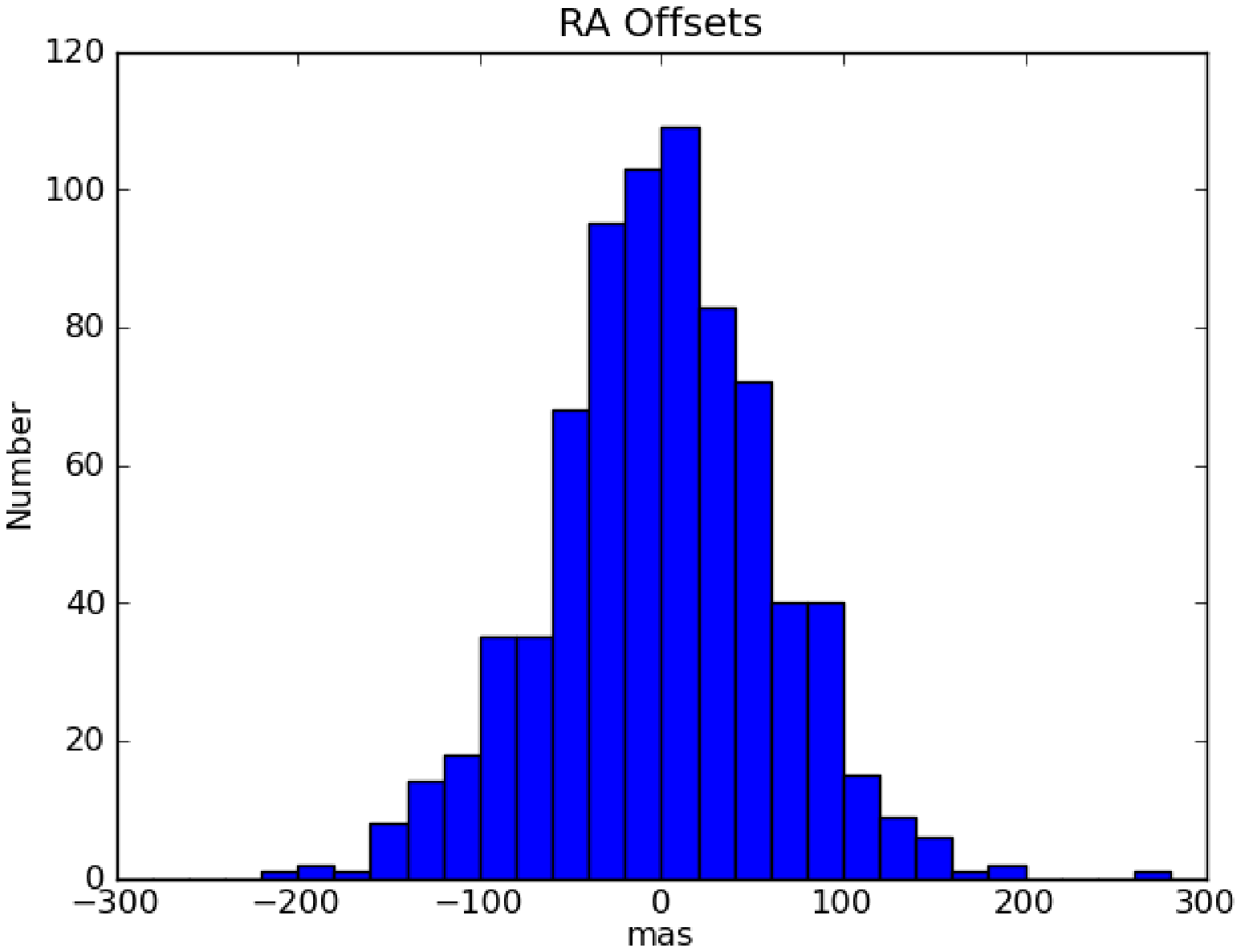}{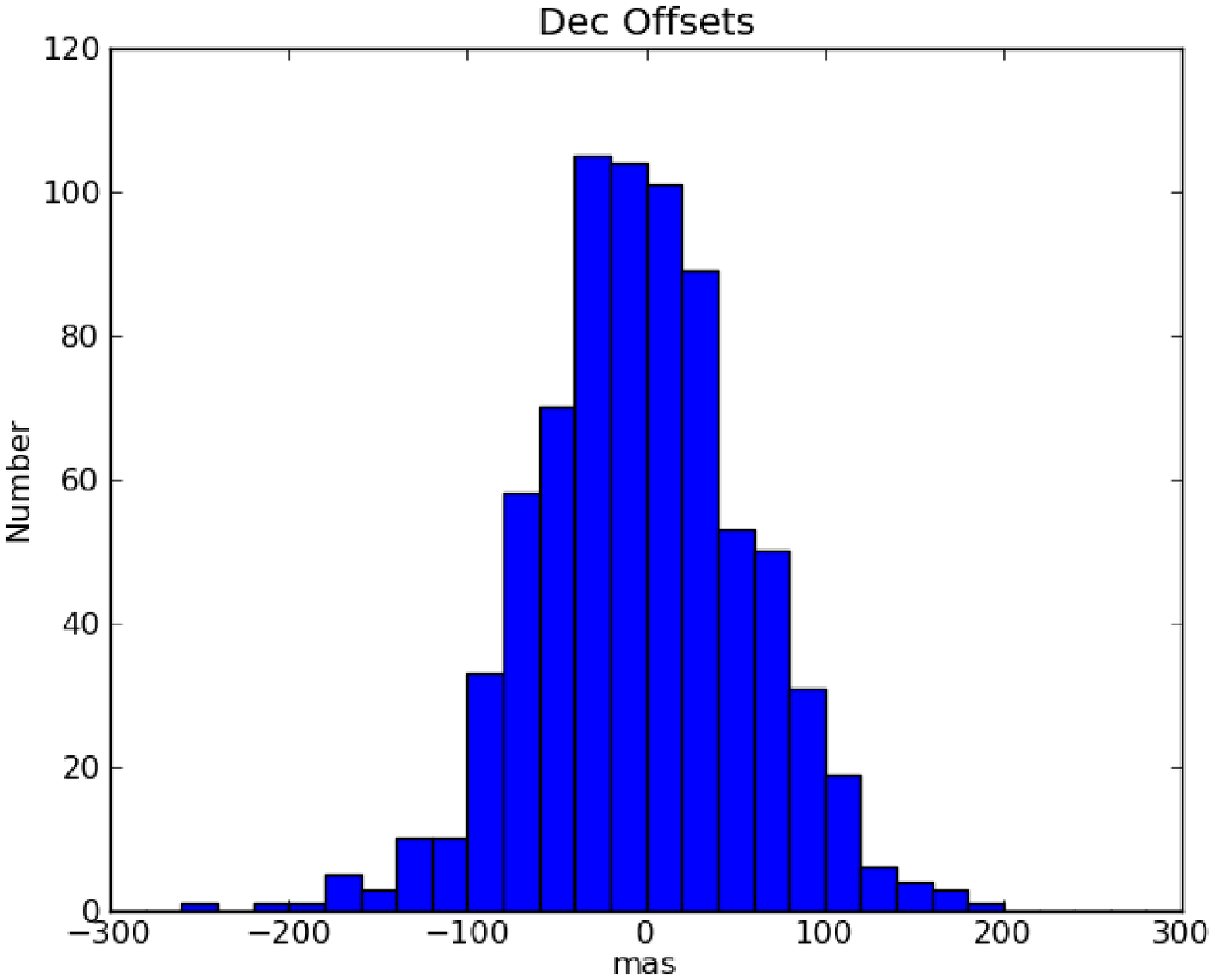}
\caption{
Position residuals in our GAS solution calculated for a subsample of 758 validation quasars.
} \label{gsimshift}
\end{figure}

We ran the solution with this reduced quasar catalog and solved for the `validation' quasars along with the grid stars. We show the histograms of the position residuals in Figure~\ref{gsimshift}. We estimate both systematic errors (mean and error of the mean), and random errors (standard deviation and we also show the normal quantiles in parentheses): the systematic errors are -0.75$\pm$2.25~mas for RA and 3.04$\pm$2.17~mas for Dec, while the random errors are 62.1~mas (-56.0, 57.2) for RA and 59.8~mas (-60.3, 57.1) for Dec. In all the validation measurements that follow we calculate the errors in the same format. The normal quantiles are equal to the standard deviation if the distribution is normal, and therefore provides aditional information about the error distribution. While the systematic errors are not significant, the random errors are almost six times larger than expected from simulations (see \S \ref{sim}).

\begin{figure}
\epsscale{0.9}
\plottwo{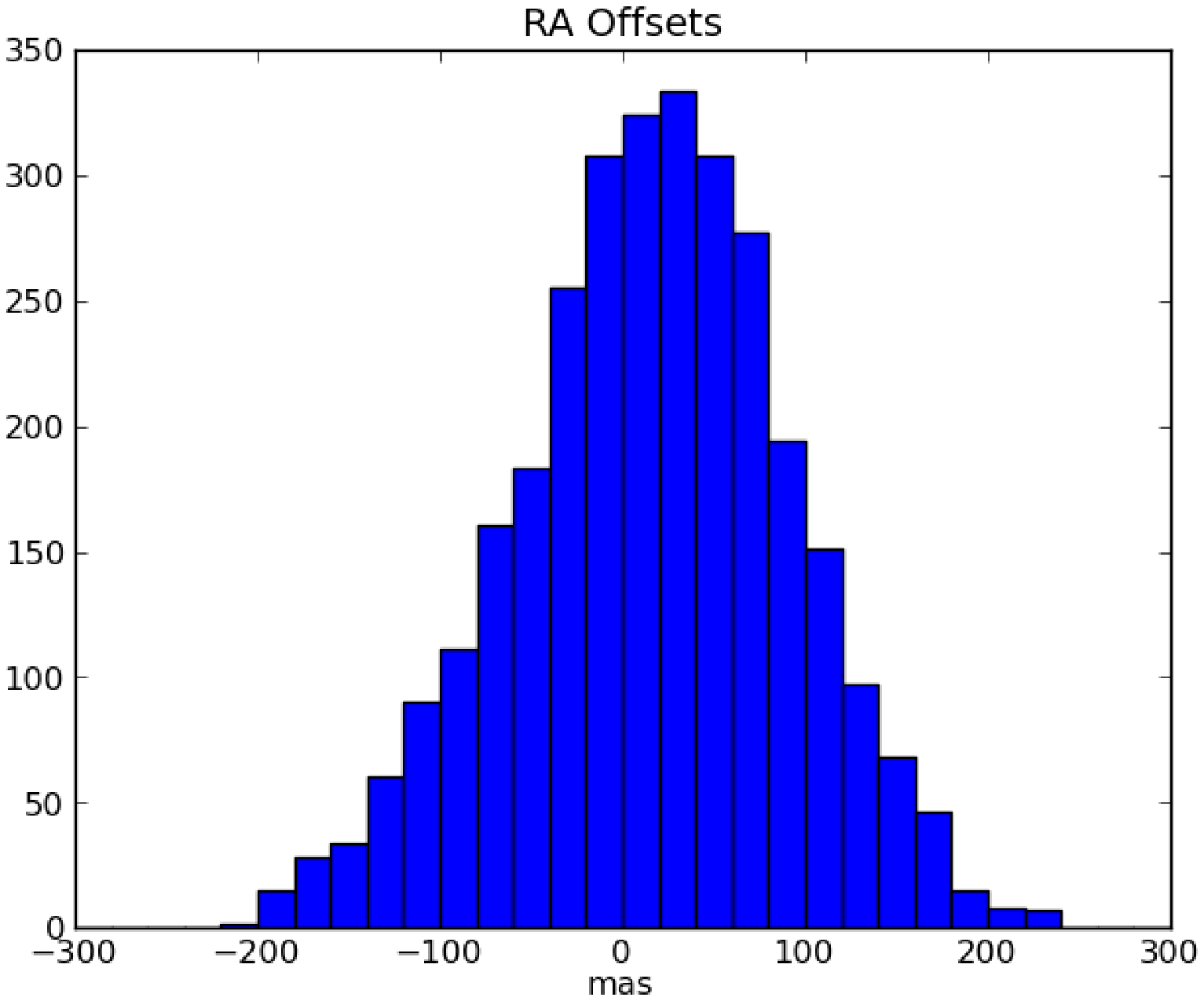}{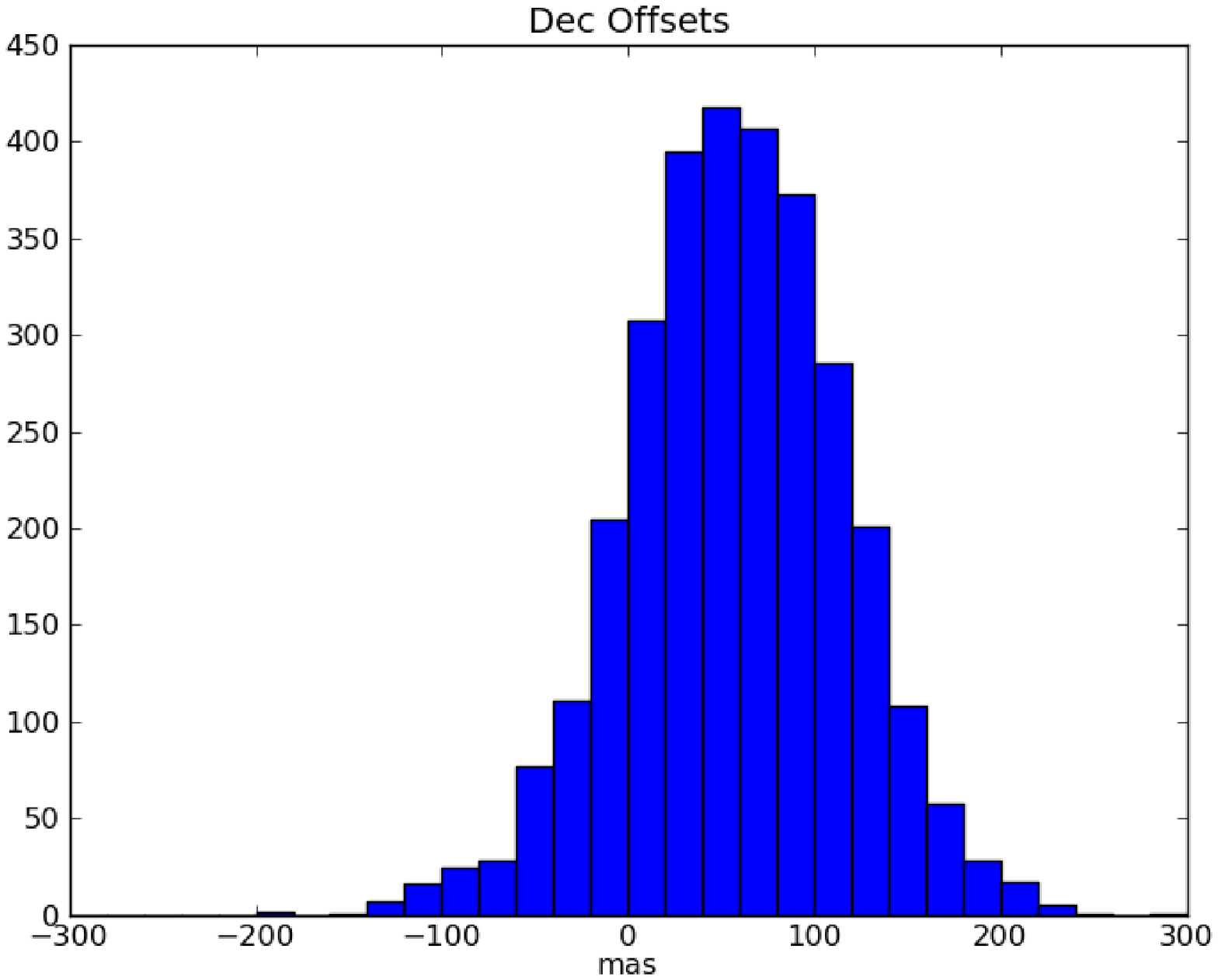}
\caption{
Original positional offsets in the PS1 pipeline solution for the quasars catalog.
} \label{ocarsshift}
\end{figure}

We compared these results with the original sky position results in the PS1 pipeline. In this case, we use all the quasars in the catalog to calculate offsets from the catalog positions. In Figure~\ref{ocarsshift} we show histograms for the positional offsets in the original PS1 pipeline solution. The systematic errors are: 15.3$\pm$1.36~mas for RA and 56.8$\pm$1.05~mas  in Dec. The random errors are 75.3~mas (-61.2, 90.1) for RA and 58.3~mas (0.9, 114.3) in Dec. There is a significant systematic shift in RA and a very large shift in Dec, which was explained in the introduction and \S \ref{icrfcat} as systematic offsets inherited from 2MASS. In those sections we investigated both the systematic and random offsets between the PS1 position and the radio position of the quasars in our catalog (see also Fig. \ref{hist.fig}). We have used different methods to filter out the objects which do not qualify as RORFO. However large offsets still exist both in the original PS1 positions and in our solution. 

\begin{figure}
\epsscale{1.0}
\plottwo{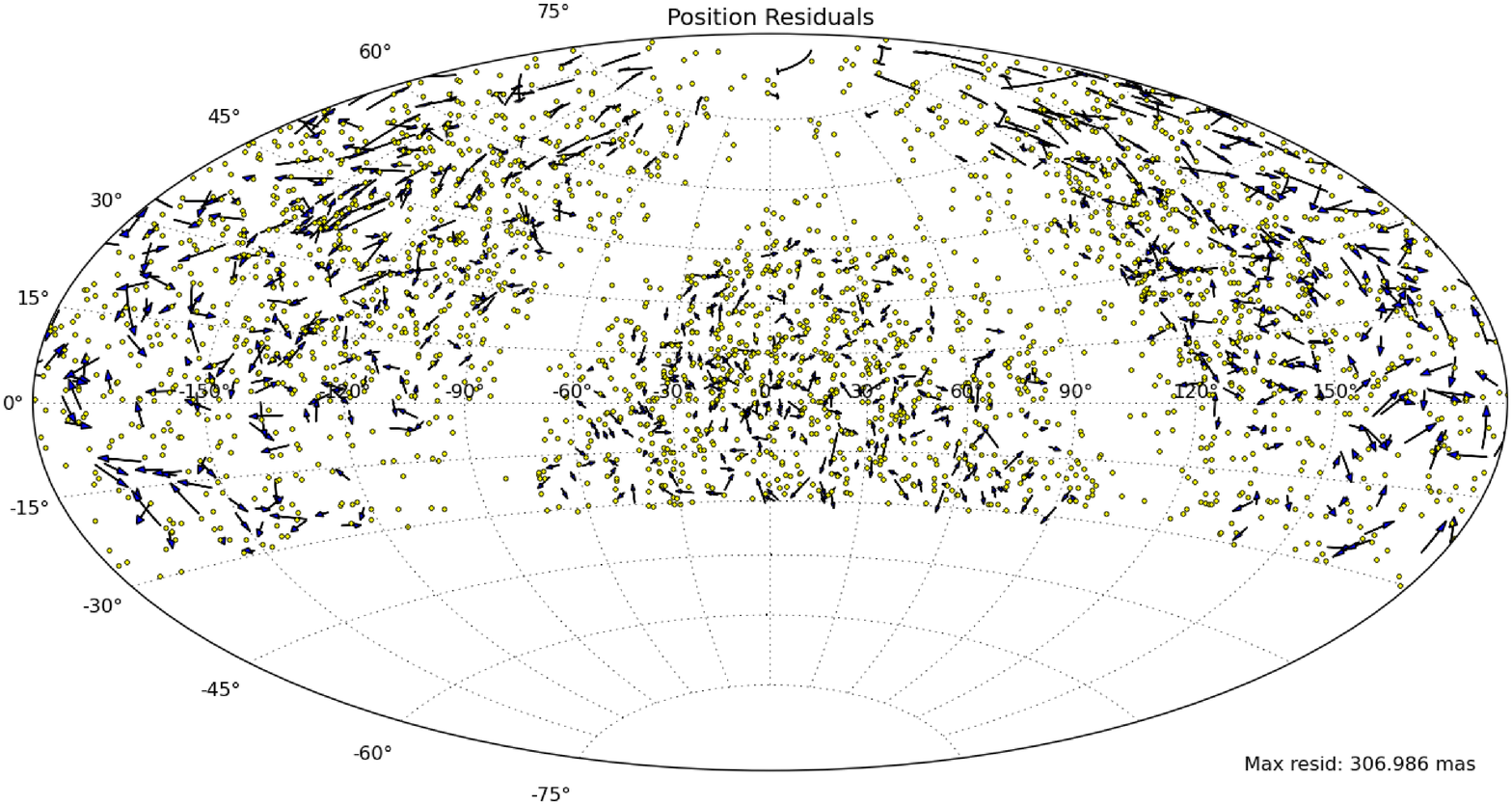}{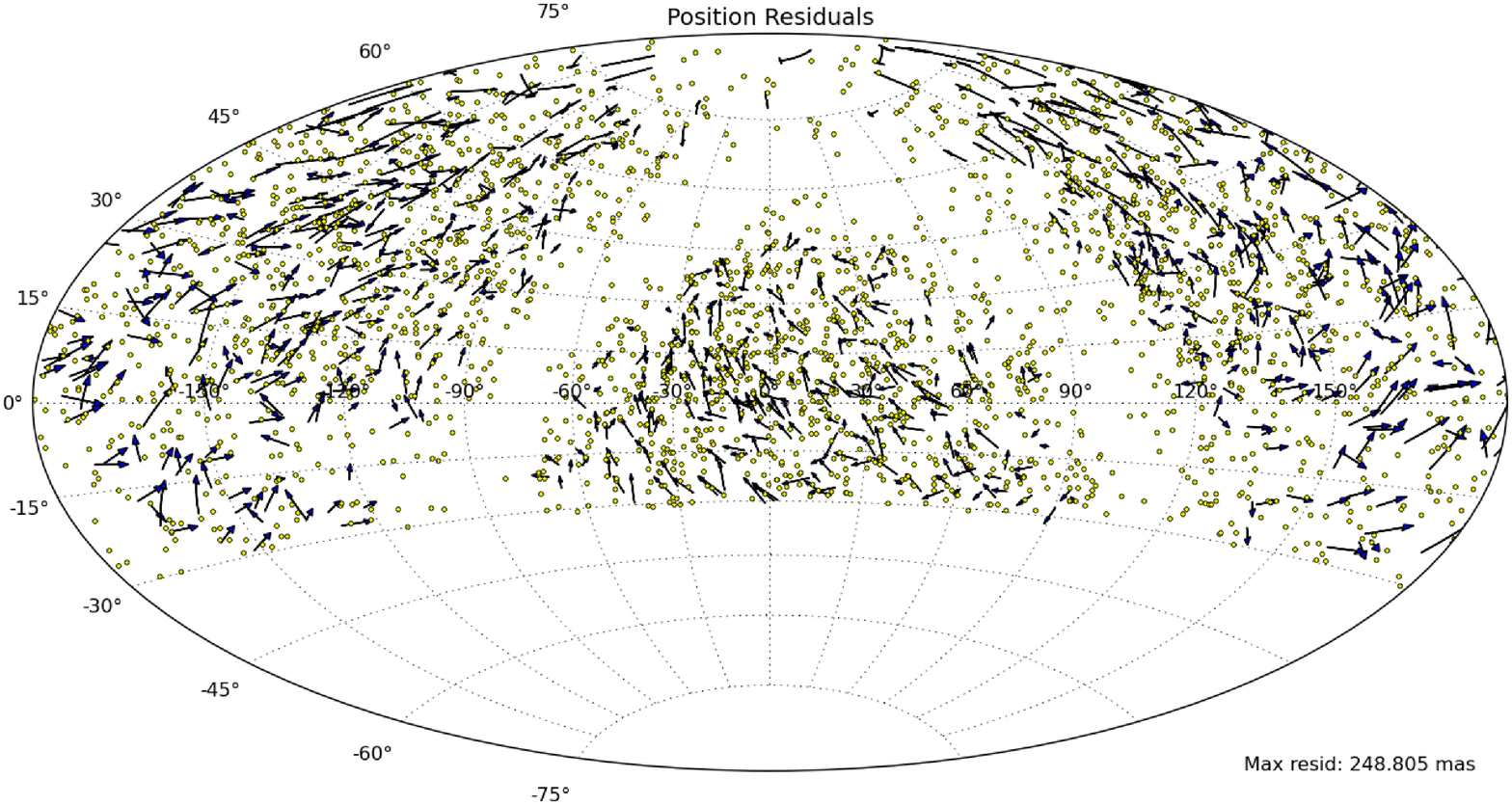}
\caption{
Position errors for the subsample of 758 validation quasars. Our solution on the left and the original PS1 pipeline solution on the right.
} \label{758pos}
\end{figure}

These results suggest that our solution removes the systematic errors correctly but does not improve significantly the coordinates on small scales. For comparison, in Figure~\ref{758pos} we show the position shifts on the sky for the subsample for 758 `validation' quasars, for our solution (left) and the original PS1 solution (right). The large-scale zonal errors are clearly seen on the right plot.

\subsubsection{Using URAT1}
To further investigate why the positional errors are so large we compare our results with the URAT1 catalog \citep{urat}. URAT1 covers almost the same amount of sky as PS1 (north of -15 deg, the southern hemisphere observations started in the late 2015) and the data are practically simultaneous with PS1 and therefore not affected by errors in proper motion. The position errors of URAT1 are 10-20 mas and are therefore sufficiently accurate to test our PS1 results. We cross-match the grid catalog with URAT using a 1$\arcsec$ cone search, and we obtain matches for 82\% of the stars. 

\begin{figure}
\epsscale{1.0}
\plottwo{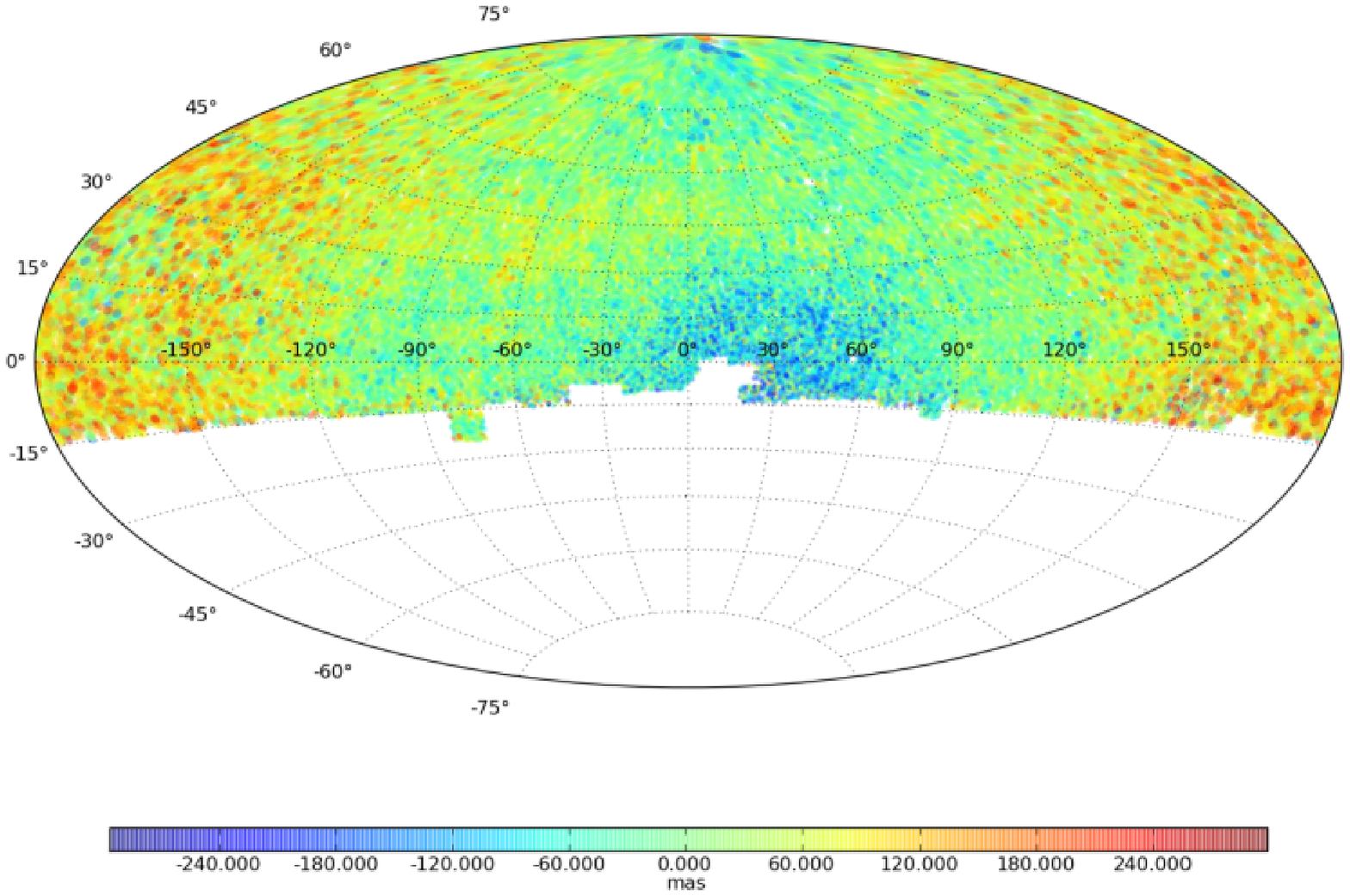}{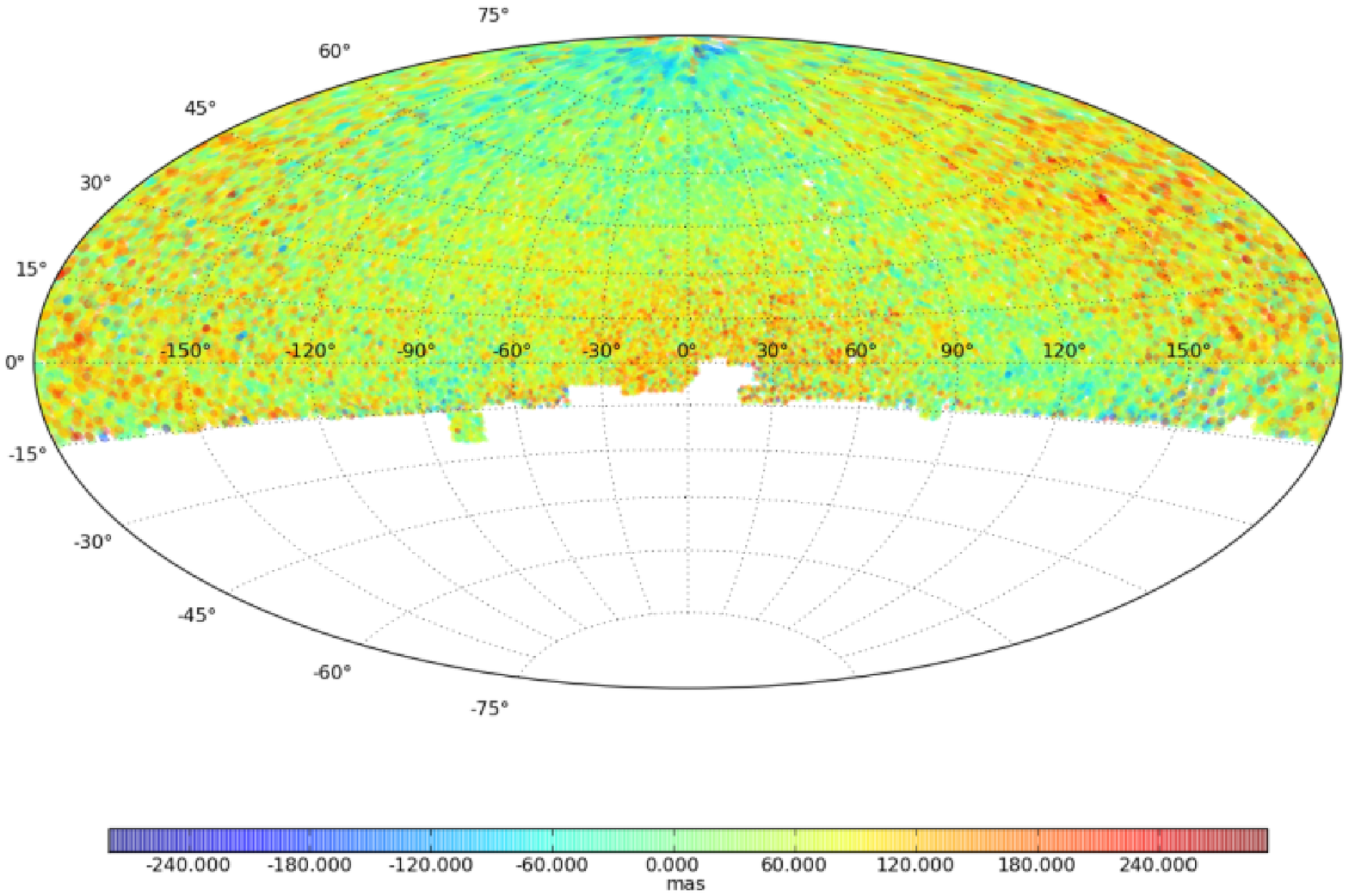}
\caption{
Positional difference between PS1 pipeline solution and URAT1; left: RA, right: Dec.
} \label{ps1-urat}
\end{figure}

First we compare the original PS1 pipeline solution (median of the original sky positions in the PS1 data) with the URAT1 positions. The differences are plotted in Figure~\ref{ps1-urat}. We notice large zonal differences in both plots: 16.1$\pm$0.10~mas in right ascension and 58.2$\pm$0.08~mas in declination. The random errors are 76.0~mas (-51.3, 84.2) in right ascension, and 64.4~mas (-0.2, 116.0) in declination. We note that these values are very similar to the offsets found above using the validation quasars $-$ confirming again that the systematic errors in PS1 are real.

\begin{figure}
\epsscale{1.0}
\plottwo{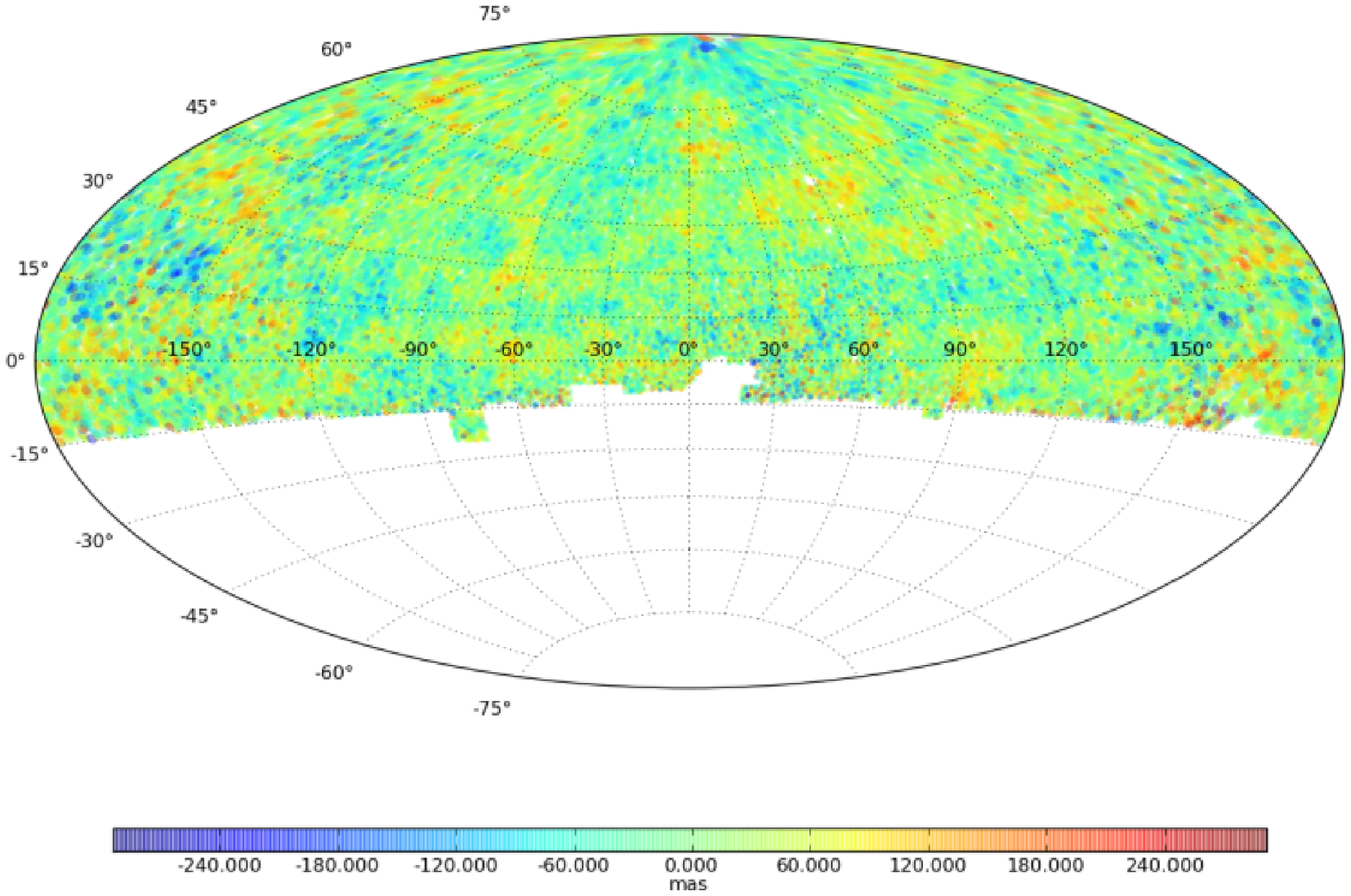}{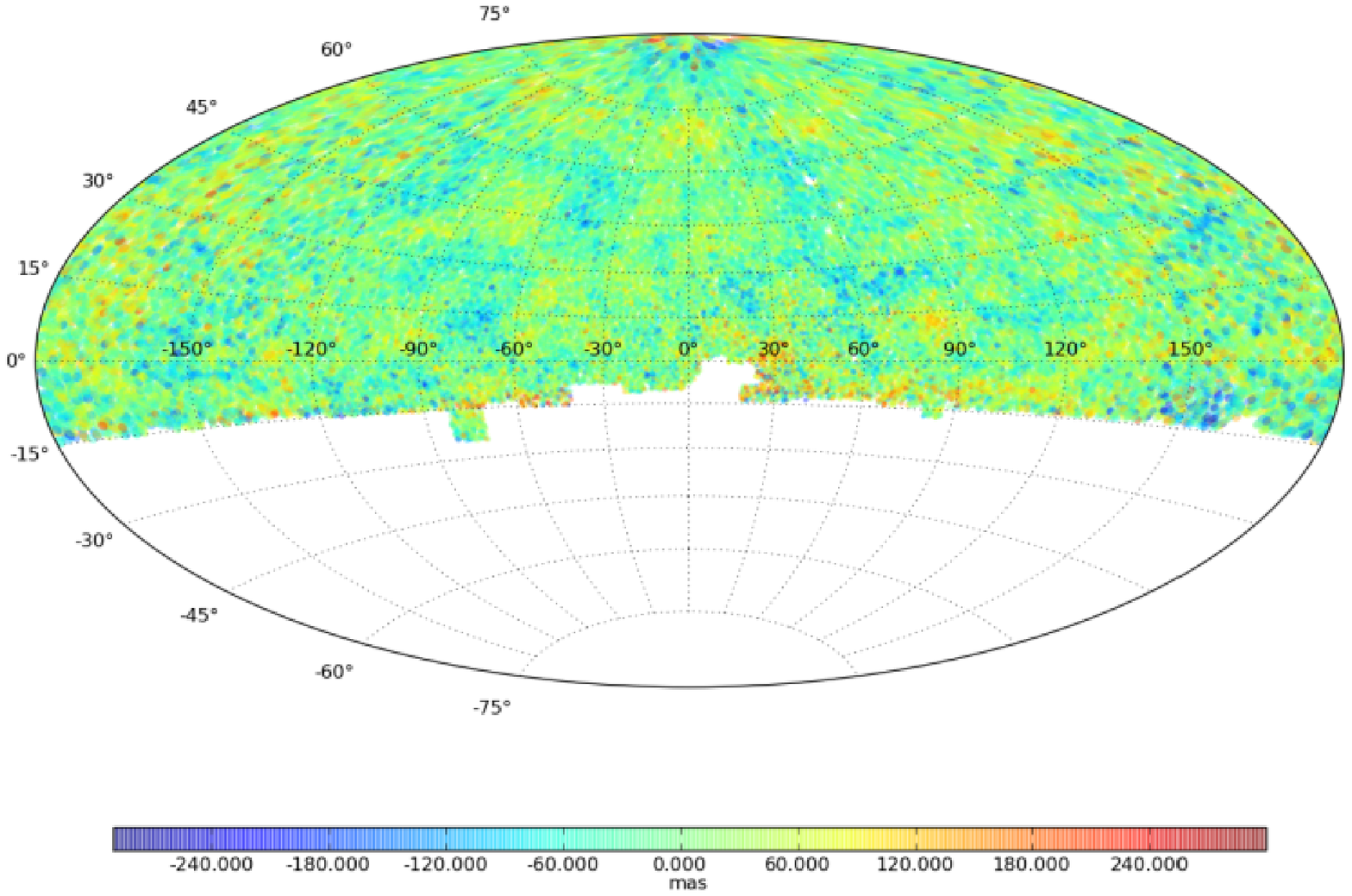}
\caption{
Positional differences between our GAS solution and URAT1; left: RA, right: Dec.
} \label{gsim-urat}
\end{figure}

\begin{figure}
\epsscale{1.0}
\plottwo{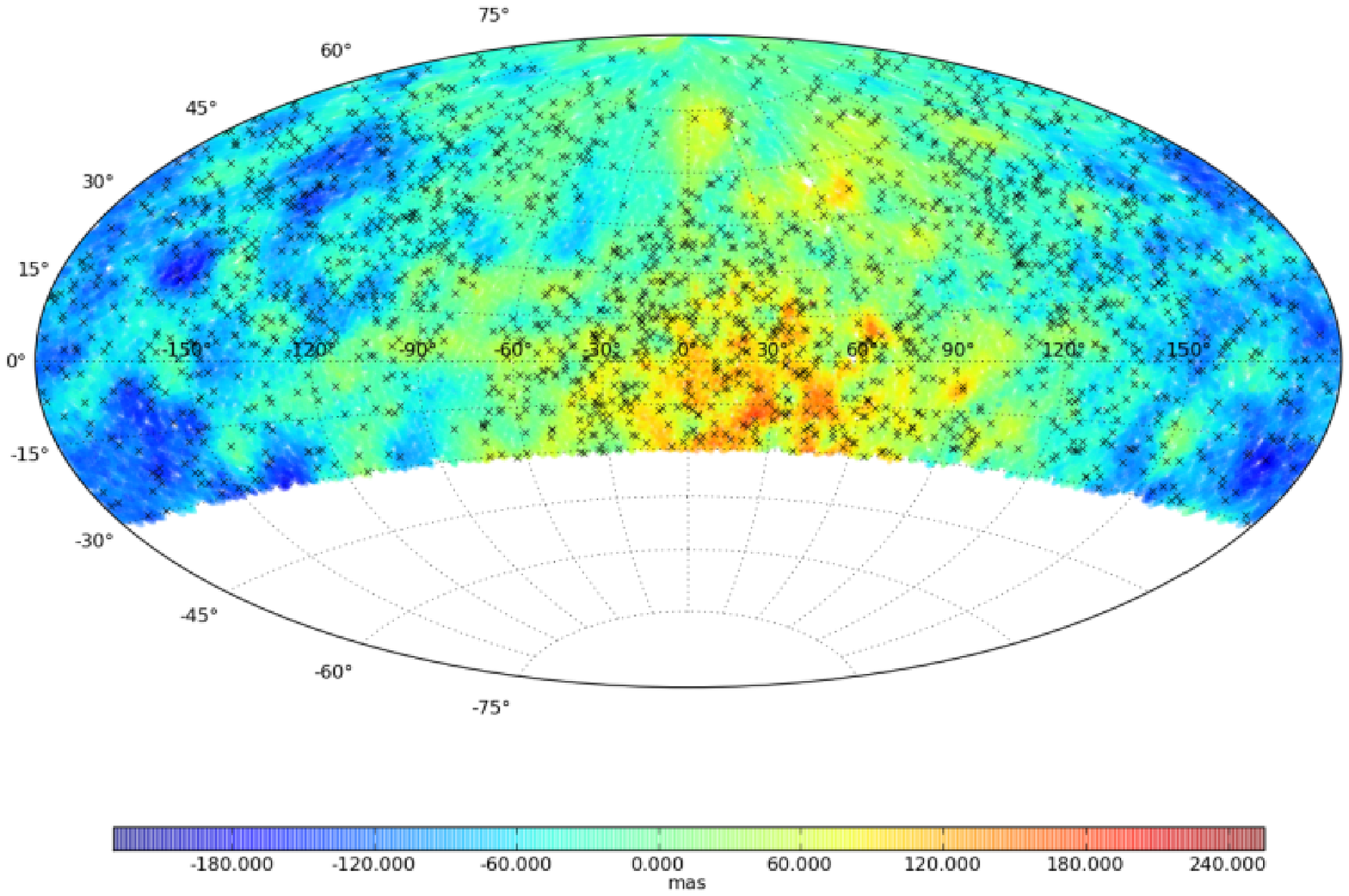}{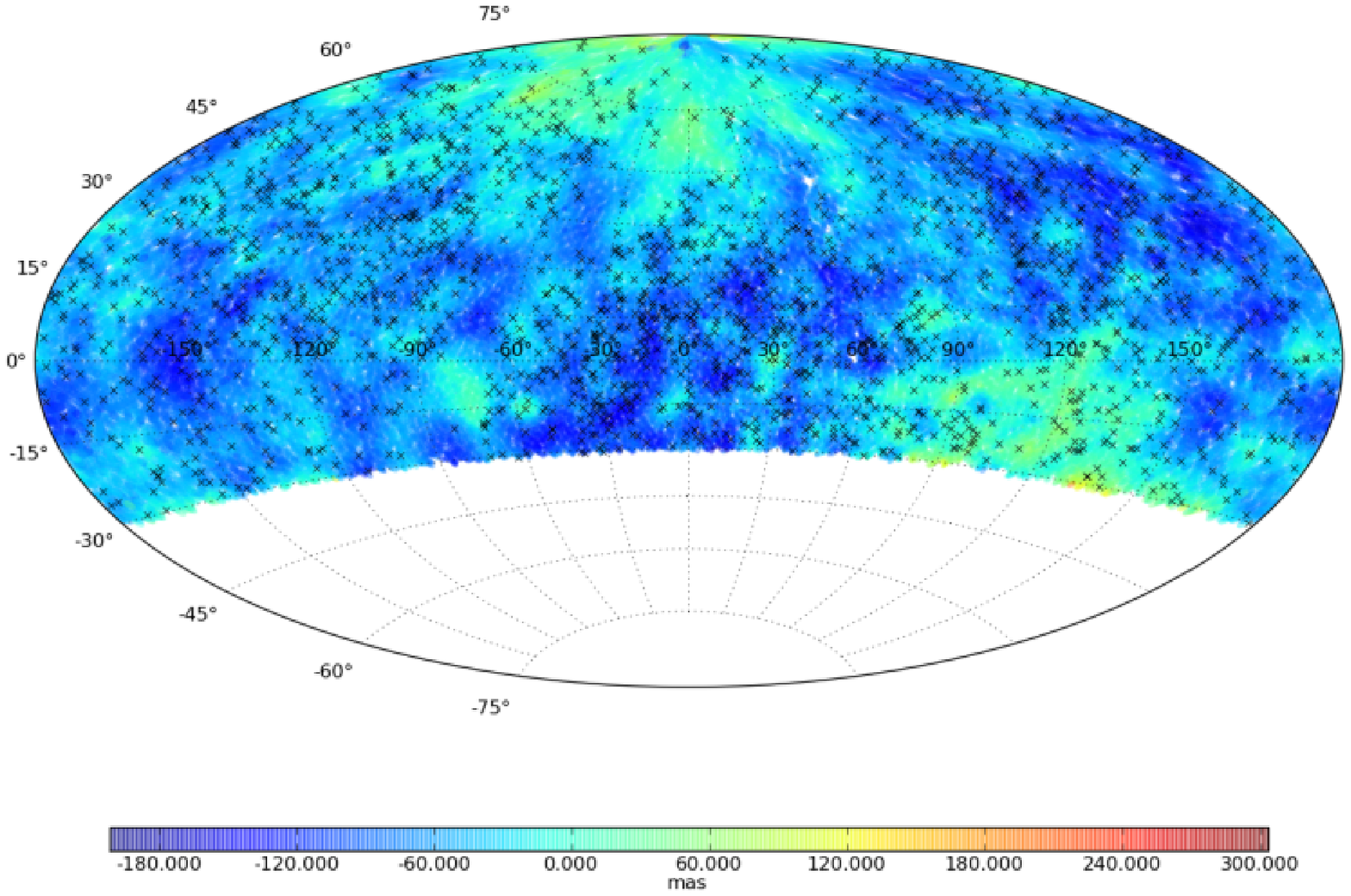}
\caption{
Positional differences between our GAS solution and the PS1 pipeline solution. left: RA, right: Dec.
} \label{gsim-ps1}
\end{figure}

Next we compare our results with URAT1, the offsets are 4.1$\pm$0.08~mas and 1.0$\pm$0.08~mas (see Figure~\ref{gsim-urat}). The random errors are 64.3~mas (-55.1, 63.0) and 62.7~mas (-56.2, 58.4). Just as we found before, the large systematic offsets (zonal errors) are now gone, but the small-scale errors did not improve significantly from the original positions in the PS1 data. Figure~\ref{gsim-ps1} shows the differences between our solution and the original PS1 pipeline positions. By comparing these plots further with Figure~\ref{ps1-urat} we see that our algorithm successfully corrects for the large-scale zonal errors but it does not improve the PS1 solution much at smaller scales.

\subsubsection{Small-scale residuals}

\begin{figure}
\epsscale{1.0}
\plottwo{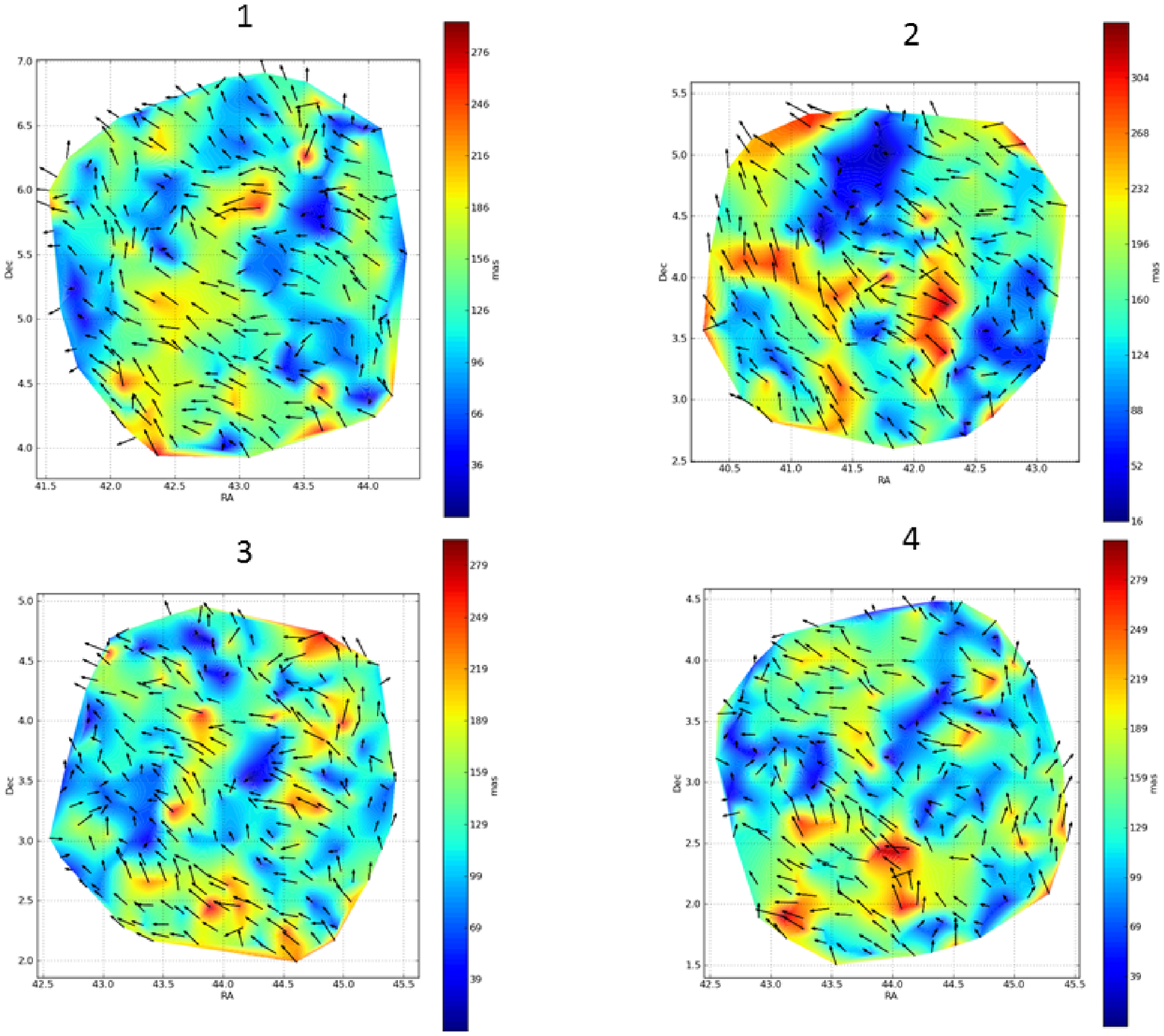}{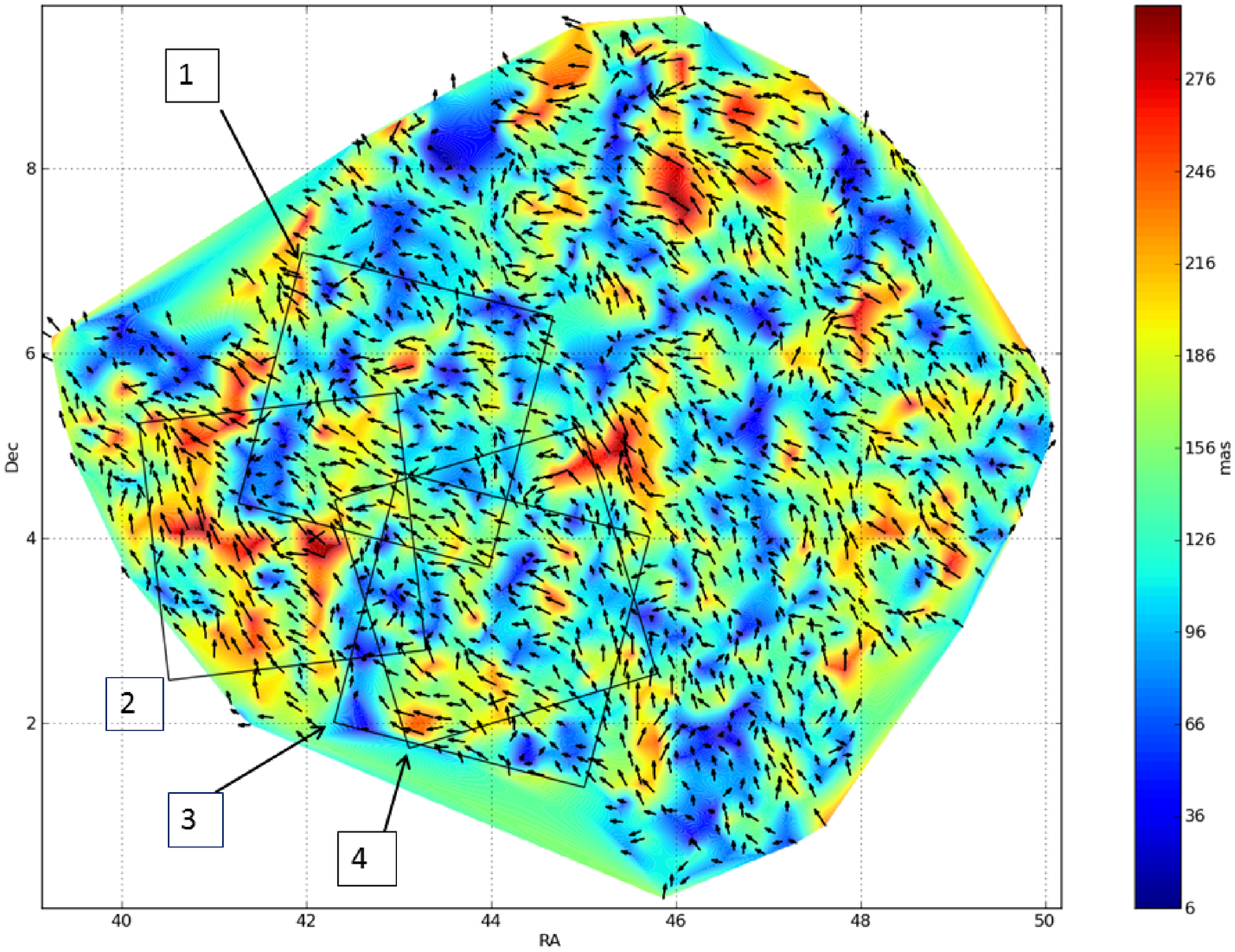}
\caption{
PS1 - URAT1 offsets at small scale using the original PS1 pipeline astrometry. The colors show absolute offset contours, with the individual offsets shown as arrows.  Left: four individual frames. Right: average PS1 positions offsets from URAT1 with the positions of the frames marked.  This plot shows the same data as Figure~\ref{ps1-urat} on a smaller scale.
} \label{tile}
\end{figure}

In Figure~\ref{tile} right, we show the same offsets as in Figure~\ref{ps1-urat}, (the offsets between URAT positions and the median positions in the original PS1 data) for a small region on the sky. This time we plot the absolute differences with arrows and we also show color coded contours. We clearly see the large scale systematic offsets; the arrows have a general direction to the left-upper corner. However, there are also significant small-scale patterns, shown by the colored contours; these are at scales of a fraction of a degree. Four individual frames are shown in the left panel and their sky positions are shown in the right panel. We notice that the patterns of individual frames on the left match the patterns seen on the sky on the right. This can only happen if this small-scale pattern exists in the entire PS1 data set. If the small scale errors were caused by distortions, or other frame-related errors (such as small shifts of the individual chips)  we would expect each frame to have different error patterns. While the large zonal errors were explained by the lack of proper motion correction, the small scale errors cannot have the same explanation.

\begin{figure}
\epsscale{0.5}
\plotone{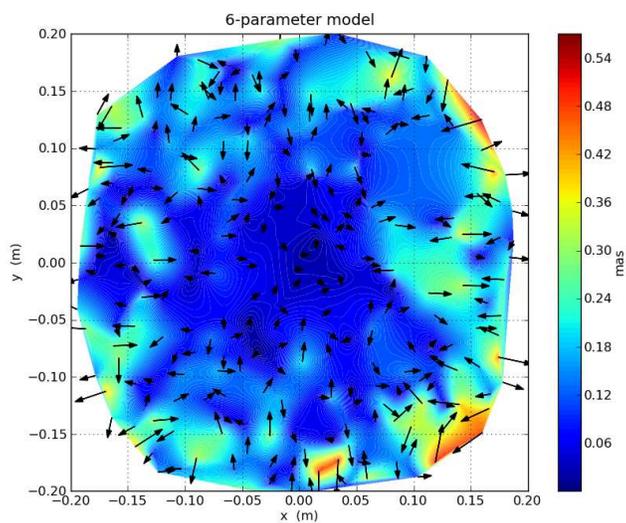}
\caption{
Residuals on one frame after fitting the PS1 mosaic coordinates to the PS1 pipeline sky coordinates.
} \label{frame}
\end{figure}

All these comparisons suggest that sky correlated errors in the PS1 data exist on both large and small scales. While the former are removed by our method, the latter are not.  This implies that the small-scale errors are also present in the mosaic coordinates used for the analysis. We tested this hypothesis by comparing the mosaic coordinates with the sky coordinates from a single frame. We found that they are related by a simple projection using a first order plate model. After correction, the differences between these coordinates are usually a fraction of a mas (see Figure~\ref{frame}). In other words, we believe the sky-correlated errors in the PS1 data propagate into our solution through the mosaic coordinates and strongly reduce our ability to improve the astrometry at smaller scales.

\subsection{Proper Motion Validation}

To validate the proper motions we are using the same methods we used for positions, namely comparisons with the subsample of 758 quasars and then with the 82\% of the stars that are common with URAT1. We remind our reader that our data do not include high proper motion stars ($\gtrsim$ 100 mas~yr$^{-1}$). This is because we used a cone search radius of 1 arcsec from the UCAC4 position to get PS1 data, and the time difference between UCAC4 and PS1 observations is $\sim$10 years.

\begin{figure}
\epsscale{0.9}
\plottwo{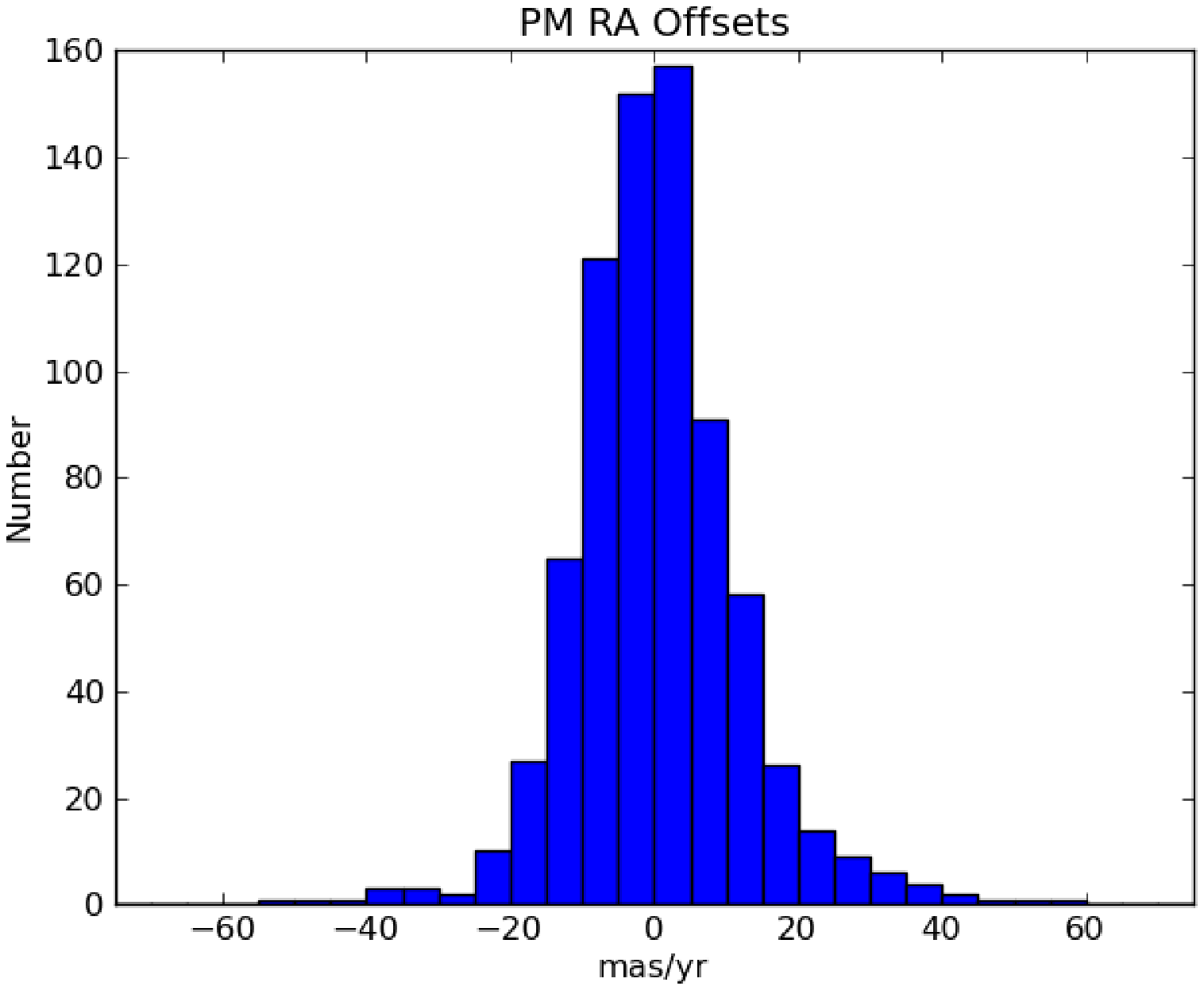}{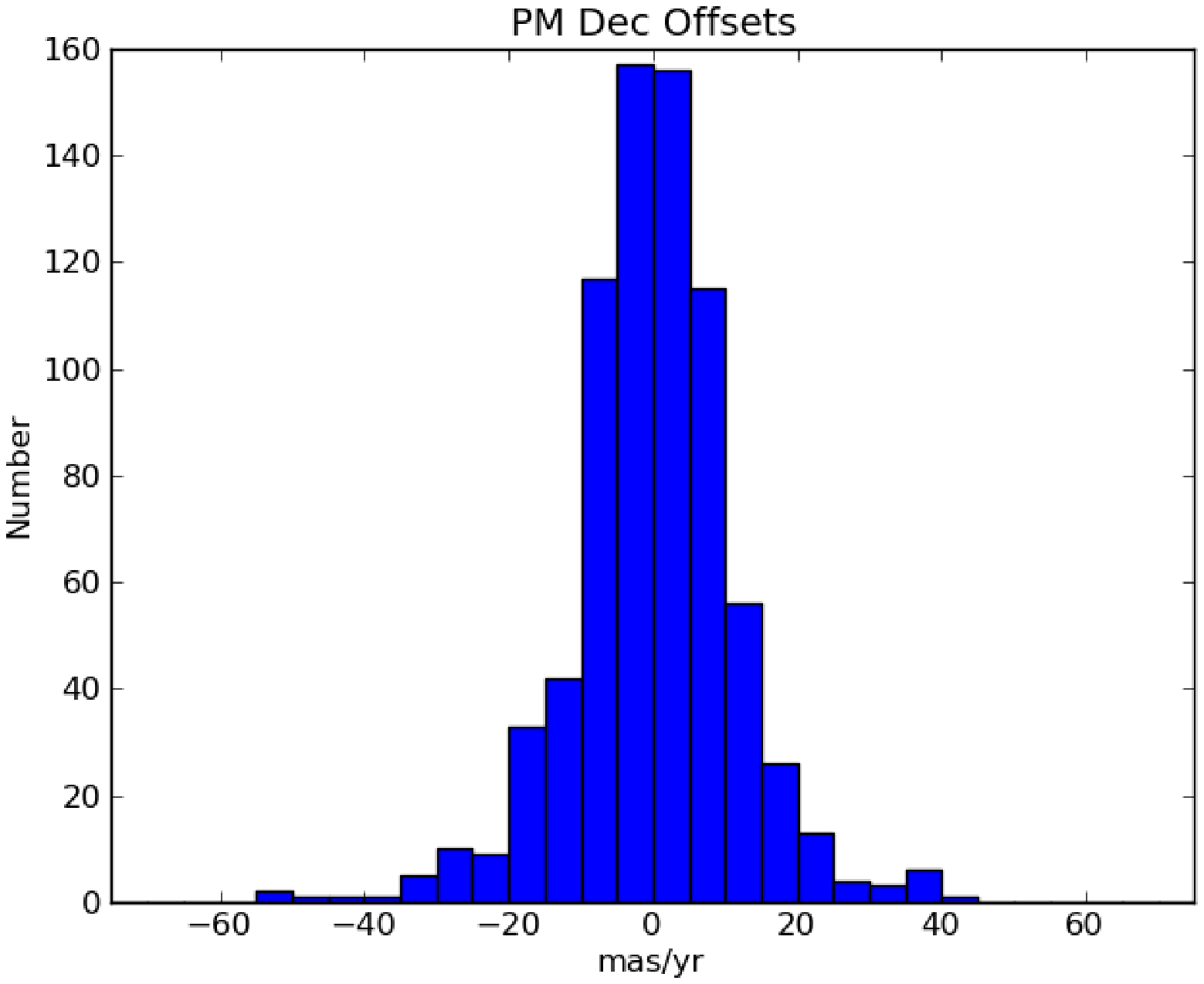}
\caption{
Proper motions residuals calculated in our global solution for a subsample of 758 validation quasars.
} \label{gsimpm}
\end{figure}

\begin{figure}
\epsscale{0.8}
\plotone{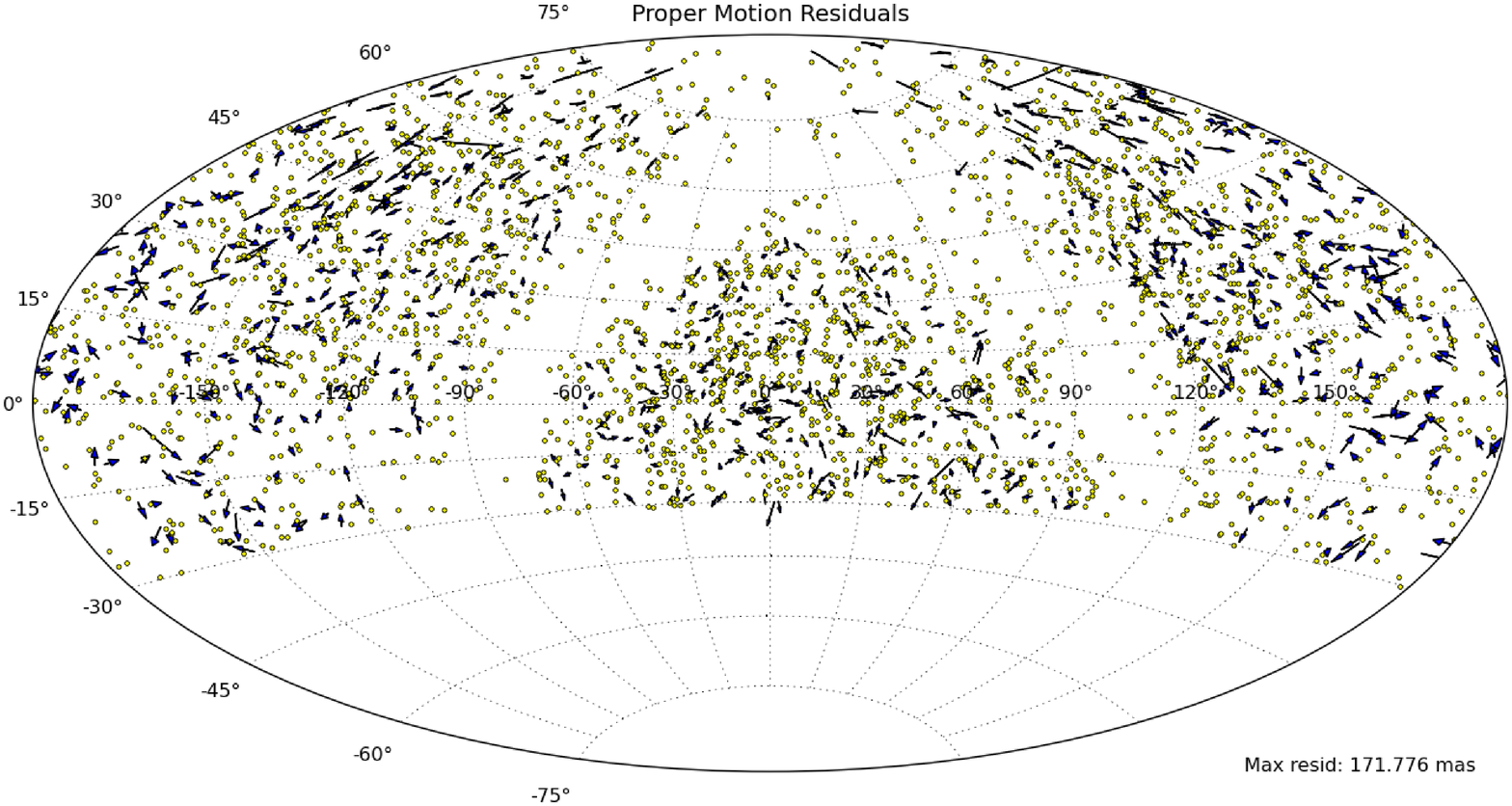}
\caption{
Proper motion errors in our solution for the subsample of 758 validation quasars.
} \label{758pm}
\end{figure}

The quasars have zero proper motion so the offsets we calculate for them represent the errors in our solution. We obtain much better results compared to position (see Figure~\ref{gsimpm}). The systematic errors are 0.55$\pm$0.49~mas~yr$^{-1}$ in RA and -0.07$\pm$0.41~mas~yr$^{-1}$ in Dec. The random errors are 13.6 mas~yr$^{-1}$ (-9.5, 10.4) in RA and 11.3~mas~yr$^{-1}$ (-9.0, 9.7) in Dec. These values are consistent with what we expected from simulations. In Figure~\ref{758pm} we show the distribution of these errors on the sky. There are no large-scale zonal errors.

\begin{figure}
\epsscale{1.0}
\plottwo{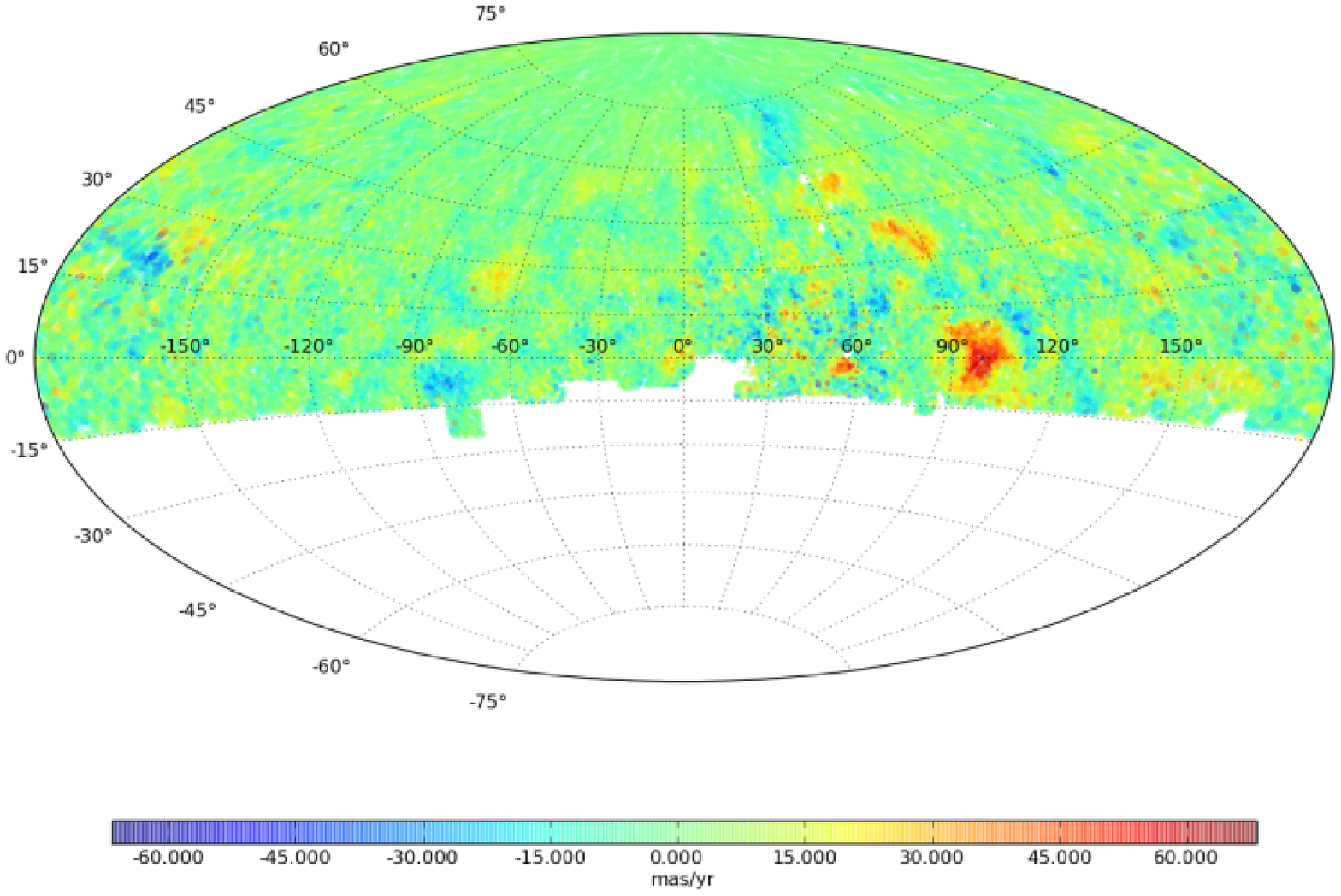}{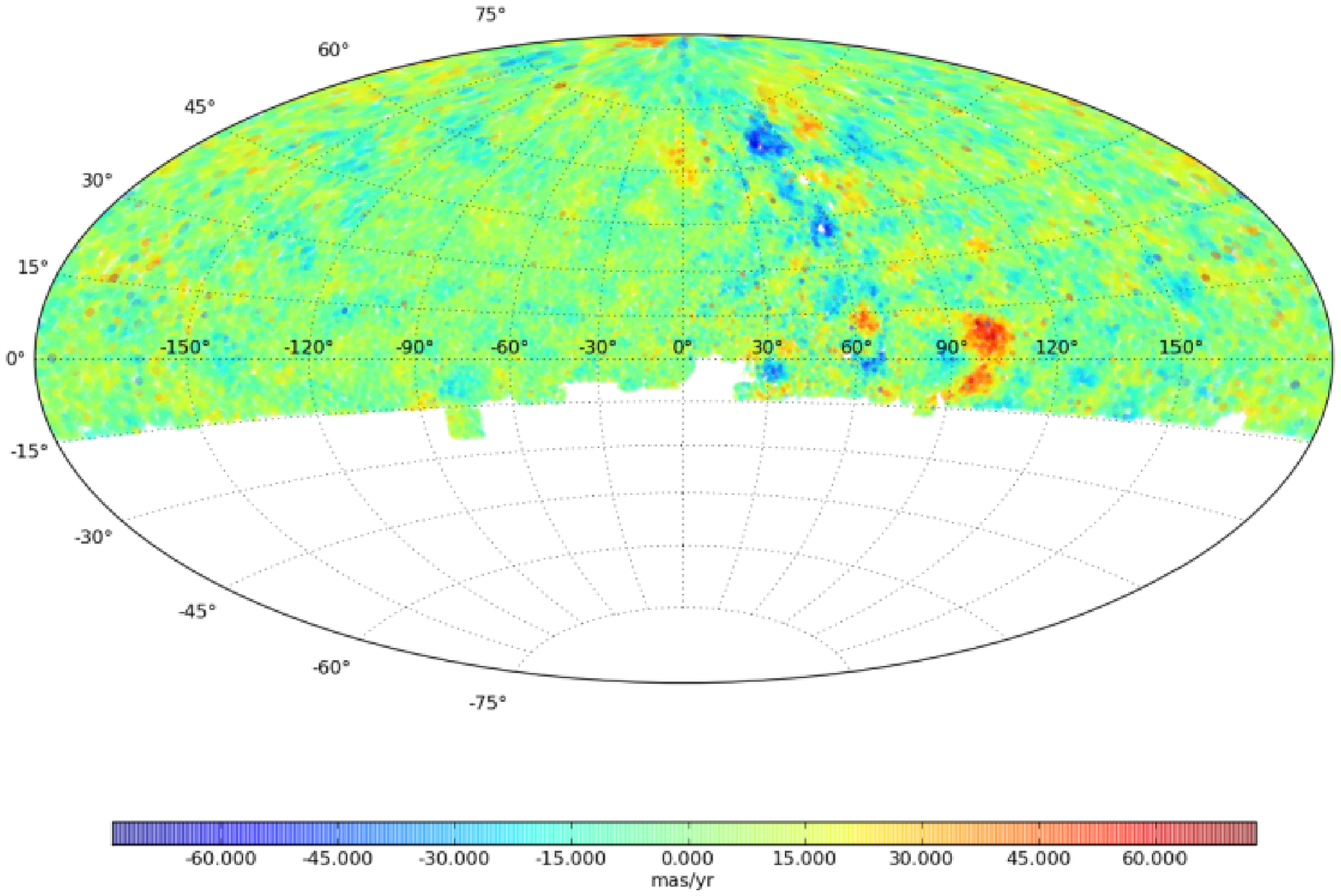}
\caption{
Comparison of our proper motion results with URAT1. left: RA, right: Dec.
} \label{gsim-urat-pm}
\end{figure}

In Figure~\ref{gsim-urat-pm} we show the comparison with URAT1 proper motions. The systematic differences are 1.2$\pm$0.014~mas~yr$^{-1}$ for RA and 1.1$\pm$0.016~mas~yr$^{-1}$ (-10.0, 12.1) for Dec. The random errors are 11.1 mas~yr$^{-1}$ (-8.0, 10.3) for RA and 12.7 mas~yr$^{-1}$ (-10.0, 12.1) for Dec. These errors are consistent with those obtained using the quasars and also with the simulations. The large-scale sky correlated errors we found in the mosaic coordinates did not affect the proper motions.  There are however some sky areas in Figure~\ref{gsim-urat-pm} which show larger errors, in particular on the right side and especially close to the Galactic plane. They seem to be correlated with areas with poor time coverage (see Figure~\ref{overtime}, right) and low density of quasars (Figure~\ref{ocars}), and are probably caused by weak constraints in specific sky areas.

\section{CONCLUSION}

We presented a rigorous method for calculating absolute astrometry for full sky (or large coverage) missions. The GAS approach provides absolute astrometry because it does not rely on previous measurements. Instead, the solution is tied directly to a relatively small number of QSOs or distant AGNs and the solution is solved simultaneously for all the grid stars in a very large, least-squares solution. However, BA is computationally intensive and requires a very careful filtering of the data both for the reference sources and the grid stars. 

We applied this method to PS1 data, which covers 3$\pi$ of the northern sky and relies on the 2MASS catalog for astrometry. We created a reference sample of 3076 QSOs and AGNs based on the OCARS catalog, and a grid catalog of 750,000 stars based on the UCAC4 catalog. Both were carefully selected to obtain uniform coverage on the sky. In particular for the reference stars a strict selection process was applied to ensure a sample of RORFO with high accuracy optical positions based on VLBI radio positions. When available we used images to investigate large offsets observed between the optical and radio positions, especially for brighter optical sources. These are most likely real offsets, and include extended, double and perturbed galaxies. We used the URAT1 catalog to remove an additional number of RORFO candidates with large radio-optical offsets in both PS1 and URAT1.

We applied several filters to remove data which can introduce systematic errors or produce an ill-conditioned system. In our solution we used the so-called mosaic coordinates which are reconstructed from the individual chip coordinates. We validate our results using simulations, quasars, and the URAT1 catalog. We show that the systematic errors in the original PS1 astrometric solution (mainly caused by using the 2MASS catalog as reference) are easily removed using our approach. However, the absolute positional errors are of the order of 60~mas, almost 6 times larger than expected from simulations, while the proper motion errors are consistent with the simulations ($\sim$10 mas yr$^{-1}$).

A comparison with the URAT1 catalog reveals that there are systematic errors on the sky in the mosaic coordinates on small scales (a fraction of the full detector). Such errors cannot be corrected using the GAS method and must be removed through calibration. The proper motion solution is not affected by these errors and shows a good correlation with the URAT1 solution. 

{\it Gaia} mission will soon provide high accuracy positions for aproximately one billion stars. These can be used to perform relative astrometry and correct both the large and small-scale errors in PS1 data. Since PS1 goes deeper than {\it Gaia} by at least a magnitude this could be of interest in cases when fainter objects are observed or higher density of sources is required. The first release of {\it Gaia} will only contain limited proper motions \citep{mic14} and therefore using {\it Gaia} stars as reference might introduce some zonal errors similar to the existing biases described in the introduction (the average epoch difference between {\it Gaia} and PS1 is about 4 years). However these can be corrected using the proper motions we calculated.

\acknowledgments

We thank the anonymous referee for carefully reading the manuscript. We also acknowledge help with the software and ideas from several coworkers not listed as authors. The PanSTARRS1 Surveys (PS1) have been made possible through contributions of the Institute for Astronomy, the University of Hawaii, the PanSTARRS Project Office, the MaxPlanck Society and its participating institutes, the Max Planck Institute for Astronomy, Heidelberg and the Max Planck Institute for Extraterrestrial Physics, Garching, The Johns Hopkins University, Durham University, the University of Edinburgh, Queen's University Belfast, the Harvard Smithsonian Center for Astrophysics, the Las Cumbres Observatory Global Telescope Network Incorporated, the National Central University of Taiwan, the Space Telescope Science Institute, the National Aeronautics and Space Administration under Grant No. NNX08AR22G issued through the Planetary Science Division of the NASA Science Mission Directorate, the National Science Foundation under Grant No. AST1238877, the University of Maryland, and Eotvos Lorand University (ELTE) and the Los Alamos National Laboratory.

\appendix

\section{General Equations for the global solution} \label{bozomath}

General Eichhorn equations are linearized by taking the first order derivatives of a Taylor expansion and calculating the Jacobian coefficients. For each star the observed minus calculated positions on a frame are written in terms of the sky coordinate ($\alpha$, $\delta$, $\mu_{\alpha}$, $\mu_{\delta}$, $\pi$) shifts and plate coordinate (a, b, c, etc) shifts:

\begin{equation}\label{eichhorn}
\begin{split}
\begin{bmatrix}
\Delta x \\
\Delta y
\end{bmatrix}
& =
\frac{\partial (x,y)}{\partial (\alpha, \delta)}\vert_0
\left(
\begin{bmatrix}
\Delta \alpha cos(\delta) \\
\Delta \delta
\end{bmatrix}
+
(t - to)
\begin{bmatrix}
\Delta \mu_{\alpha} cos(\delta) \\
\Delta \mu_{\delta}
\end{bmatrix}
+
\begin{bmatrix}
F_{\alpha} \\
F_{\delta}
\end{bmatrix}
\begin{bmatrix}
\Delta \pi cos(\delta) \\
\Delta \pi
\end{bmatrix}
\right) \\
& +
\frac{\partial (x,y)}{\partial (a,b,c,...)}\vert_0
\begin{bmatrix}
\Delta a \\
\Delta b \\
\Delta c \\
... \\
\end{bmatrix}
\end{split}
\end{equation}

where F$_{\alpha}$ and F$_{\delta}$ are the parallax factors. The Jacobians are calculated at the assumed zero values of the parameters. The time difference $t-t_0$ is calculated between the observation time and the catalog time (in our case 2012.6 the median epoch for the PS1 data).

The normal coordinates ($\xi$, $\eta$ in units of the focal length, F$_l$) will be used as intermediate between the sky and plate coordinates. They are also called tangential coordinates because they represent coordinates in the tangential plane of the gnomonic projection. The normal coordinates do not enter the equations except as auxiliary parameters. They also help define the plate model and calibration parameters.

The most general {\bf first order} plate model has six parameters:

\begin{equation}
\begin{split}
x& = c_1 \xi + c_2 \eta + c_3 \\
y& = c_4 \xi + c_5 \eta + c_6
\end{split}
\end{equation}

The 4-parameters model includes shifts (a and b), rotation angle (c)
and the scale (d):

\begin{equation}
\begin{split}
x& = a + (\xi cos(c) - \eta sin(c)) d \\
y& = b + (\xi sin(c) + \eta cos(c)) d
\end{split}
\end{equation}

If the scale is constant at the nominal value of the focal length (F$_l$):

\begin{equation}
\begin{split}
x& = a + (\xi cos(c) - \eta sin(c)) F_l \\
y& = b + (\xi sin(c) + \eta cos(c)) F_l
\end{split}
\end{equation}

Let's calculate the coefficients for these simple 3 and 4-parameter models. The Eichhorn method calculates small shifts from the assumed values ($x_0, y_0, \xi_0, \eta_0, c_0, a_0, b_0, d_0$). For each star we should have a reasonable estimate for the sky position, and the same for the plate positions. Using these, we calculate first the assumed normal coordinates, $\xi_0$, $\eta_0$ using the gnomonic projection (equation \eqref{gnomonic}). Then we use the plate model and the assumed plate parameters ($a_0 = b_0 = 0, c_0, d_0 = F_l$)
to calculate the assumed position on the plate, x$_0$, y$_0$.

The first Jacobian coefficients can be calculated in two steps:

\begin{equation}
\frac{\partial (x,y)}{\partial (\alpha, \delta)}\vert_0
=
\frac{\partial (x,y)}{\partial (\xi, \eta)}\vert_0
\frac{\partial (\xi, \eta)}{\partial (\alpha, \delta)}\vert_0
\end{equation}

The first part has a simple form depending on the plate model, which in our simple case is just the rotation matrix (for both 3 and 4-parameters models):

\begin{equation}
\frac{\partial (x,y)}{\partial (\xi, \eta)}\vert_0
=
F_l
\begin{bmatrix}
cos(c_0) & -sin(c_0) \\
sin(c_0) & cos(c_0)
\end{bmatrix}
\end{equation}

The second part is not so simple, since the normal coordinates are given by:

\begin{equation}\label{gnomonic}
\begin{split}
\frac{\xi}{F_l}& = \frac{cos(\delta) sin(\alpha - \alpha_0)}
{sin(\delta) sin(\delta_0) + cos(\delta) cos(\delta_0) cos(\alpha - \alpha_0)} \\
\frac{\eta}{F_l}& = \frac{sin(\delta) cos(\delta_0) - cos(\delta) sin(\delta_0) cos(\alpha - \alpha_0)}
{sin(\delta) sin(\delta_0) + cos(\delta) cos(\delta_0) cos(\alpha - \alpha_0)}
\end{split}
\end{equation}

Where $\alpha_0$ and $\delta_0$ are the plate coordinates (center), not the assumed star coordinates! We will not calculate the factors here, but they are straightforward.

Finally, the last Jacobian coefficient in the equation is (for the 3-parameter model):

\begin{equation}
\begin{split}
x_0& = \xi_0 cos(c_0) - \eta_0 sin(c_0) \\
y_0& = \xi_0 sin(c_0) + \eta_0 cos(c_0)
\end{split}
\end{equation}

In this 3-parameter case this can be written in the simplified form:

\begin{equation}
\frac{\partial (x,y)}{\partial (a,b,c)}\vert_0
=
\begin{bmatrix}
1 & 0 & -y_0 \\
0 & 1 & x_0
\end{bmatrix}
\end{equation}

For the 4-parameter model we get a similar result:

\begin{equation}
\frac{\partial (x,y)}{\partial (a,b,c,d)}\vert_0
=
\begin{bmatrix}
1 & 0 & -y_0 & x_0/F_l \\
0 & 1 &  x_0 & y_0/F_l
\end{bmatrix}
\end{equation}

\section{The QR Elimination} \label{qr}

This section presents the method of \citet{mak05} to eliminate plate parameters.
Let's assume for simplicity we have the simplest plate model with only 3 parameters (the attitude parameters).
These are eliminated plate by plate. Only then are the small matrices for each plate assembled into a large matrix.  The columns must match for each individual star that falls on different plates.

For a given plate, the system of equations \eqref{gnomonic} can be written as:

\begin{equation}
A x + B y = r
\end{equation}

where x are the star parameters (5 per star), y the 3 plate parameters, and r the plate residuals. Since on a particular plate each star produces 2 equations, the size of the matrices A and B is very easy to calculate. If there are n stars on the plate, A is a 2n $\times$ 5n matrix, while B is 2n $\times$ 3.

To eliminate the 3 plate parameters, their coefficient matrix is factorized: B = QR. Q is a 2n $\times$ 2n orthogonal matrix and R is 2n $\times$ 3  upper-triangular, so that we can also write:

\begin{equation}
B = Q 
\begin{bmatrix}
R^+ \\
0
\end{bmatrix}
\end{equation}

where R$^+$ is a small 3 $\times$ 3 upper-triangular invertible matrix.

Among other applications, the QR factorization is used to solve least-square problems. Unfortunately, for huge problems such as this, it is not easy to perform. However, it can be used on the small plate matrices above to eliminate the plate parameters, following \citet{mak05}.

We multiply the equation above with Q$^T$ and at the same time we split the equations so that the bottom 2n-3 equations have only star unknowns:

\begin{equation}
\begin{split}
R^{+}y + A^{+}x &= r^{+} \\
A^{-}x &= r^{-},
\end{split}
\end{equation}

where A$^{+}$ and A$^{-}$ are the top 3 rows and bottom 2n-3 rows of Q$^T$A, r$^{+}$ and r$^{-}$ being similarly defined.

The small matrices A$^{-}$ are then assembled into a large sparse matrix containing only the star parameters. Similarly, the collected r$^{-}$ arrays form the right hand side of the equation. If we only solve for position and proper motion the resulting matrix will have approximately 3 million columns and 60 million rows but it will be very sparse. We actually solve the normalized matrix, which is 3 by 3 million symmetric positive definite, and also has $\sim$20 times fewer non-zero elements (NNZ ). If computer memory is an issue, the non-normalized matrix can be split into an arbitrary number of parts (rows-wise), each normalized separately and then added together. This approach is also very easy to parallelize for faster processing (embarrassingly parallel).

Once the star matrix is solved, the plate parameters can be recovered using the top equation above, because R$^{+}$ are small invertible matrices:

\begin{equation}
y = (R^{+})^{-1}(r^{+} - A^{+}x)
\end{equation}

Note that the process of elimination uses only orthogonal transformations and therefore the least square problem is not altered in any way. Moreover, this procedure can be used for any number of plate parameters.

\section{The Block Elimination}

This scheme follows \citet{vegt72} to eliminate the star parameters after the design matrix is generated.
As in Appendix B, the system of equations can be simply written as:

\begin{equation}
A x + B y = r,
\end{equation}

but this new formula now represents the full design matrix, with A and B as full matrices for star and plate parameters, respectively.

We also show how the errors are used to construct weighted equations. Let $\sigma$ be the error vector for both star and plate parameters, and G the covariance matrix formed from $\sigma^{-1}$. The normal equations are then:

\begin{equation}
\begin{bmatrix}
A^T \\
B^T
\end{bmatrix}
G
\begin{bmatrix}
A & B
\end{bmatrix}
=
\begin{bmatrix}
A^T \\
B^T
\end{bmatrix}
G r
\end{equation}

As shown above for each plate, the whole design matrix can be split and one set of parameters eliminated, this time without any factorization:

\begin{equation}
\begin{bmatrix}
A^TGA & A^TGB \\
B^TGA & B^TGB
\end{bmatrix}
=
\begin{bmatrix}
A^T \\
B^T
\end{bmatrix}
G r
\end{equation}

In this case it is convenient to eliminate the star parameters, because the normalized matrix A$^T$GA is 5 $\times$ 5 diagonal, and can be easily inverted. 

\begin{equation}
y
=
\frac{B^TGr - B^TGA (A^TGA)^{-1} A^TGr}{B^TGB - B^TGA (A^TGA)^{-1} A^TGB}
\end{equation}

Finally the star parameters are calculated by back-substituting the calculated y:

\begin{equation}
x
=
(A^TGA)^{-1}(A^TGr - A^TGBy)
\end{equation}

 \begin{table*}
 \centering
\tiny
 \caption{OCARS objects rejected as RORFO after visual inspection.}
 \label{rej.tab}
 \begin{tabular}{@{}lrrrrrl@{}}
 \hline
            &                 &      \\
   Name     &  RA J2000, h,m,s    & Dec J2000, d,m,s & redshift & mag. & Type & Other names and notes\\
 \hline
M81      & 09 55 33.1730 & $+$69 03 55.060 & $-$.0001 &  6.8V & AQ &  ICRF J095533.1$+$690355; NVSS J095533$+$690355 \\
SN1993J  & 09 55 24.7747 & $+$69 01 13.702 & 0.0000 & 12.0V & SN &  ICRF J095524.7$+$690113; SN 1993J \\
M84      & 12 25 03.7433 & $+$12 53 13.139 & 0.0034 & 10.6V & G  &  ICRF J122503.7$+$125313; 87GB 122232.6$+$131000 \\
         & 11 04 27.3139 & $+$38 12 31.799 & 0.0300 & 13.1V & AL &  ICRF J110427.3$+$381231 DEF; MRK 0421 \\
         & 02 48 14.8281 & $+$04 34 40.861 & 0.0237 & 13.0V & G  &  ICRF J024814.8$+$043440 (VCS$-$only); NGC 1101 \\
         & 13 36 08.2597 & $-$08 29 51.797 & 0.0231 & 13.0R & AB &  ICRF J133608.2$-$082951 (VCS$-$only); NGC 5232 \\
         & 16 06 16.0278 & $+$18 14 59.819 & 0.0368 & 13.8V & G  &  ICRF J160616.0$+$181459 (VCS$-$only); NGC 6061 \\
         & 22 49 54.5860 & $+$11 36 30.845 & 0.0262 & 13.5V & G  &  ICRF J224954.5$+$113630 (VCS$-$only); NGC 7385 \\
NGC0315  & 00 57 48.8833 & $+$30 21 08.811 & 0.0165 & 11.2V & G  &  ICRF J005748.8$+$302108; NGC 0315 \\
NGC1052  & 02 41 04.7985 & $-$08 15 20.751 & 0.0050 & 11.0V & AS &  ICRF J024104.7$-$081520; NVSS J024104$-$081521 \\
NGC1218  & 03 08 26.2238 & $+$04 06 39.300 & 0.0287 & 13.5V & AS &  ICRF J030826.2$+$040639; NGC 1218 \\
3C84     & 03 19 48.1600 & $+$41 30 42.104 & 0.0176 & 12.5V & AS &  ICRF J031948.1$+$413042; NGC 1275 \\
NGC2484  & 07 58 28.1081 & $+$37 47 11.807 & 0.0428 & 13.9V & G  &  ICRF J075828.1$+$374711; NGC 2484\\
NGC4261  & 12 19 23.2160 & $+$05 49 29.699 & 0.0075 & 11.4V & G  &  ICRF J121923.2$+$054929; NGC 4261 \\
3C274    & 12 30 49.4233 & $+$12 23 28.043 & 0.0043 & 10.8V & G  &  ICRF J123049.4$+$122328; 3C 274 \\
NGC5141  & 13 24 51.4411 & $+$36 22 42.772 & 0.0174 & 12.8V & G  &  ICRF J132451.4$+$362242; NGC 5141 \\
NGC7720  & 23 38 29.3832 & $+$27 01 53.258 & 0.0302 & 13.3V & AS &  ICRF J233829.3$+$270153; NGC 7720 \\
         & 09 43 19.1534 & $+$36 14 52.072 & 0.0225 & 16.4V & AQ &  ICRF J094319.1$+$361452 (VCS$-$only); NGC 2965 \\
UG03927  & 07 37 30.0869 & $+$59 41 03.194 & 0.0405 & 11.8R & AB &  ICRF J073730.0$+$594103; UGC 03927 \\
NGC3862  & 11 45 05.0090 & $+$19 36 22.741 & 0.0217 & 13.0V & G  &  ICRF J114505.0$+$193622; NGC 3862 \\
NGC3894  & 11 48 50.3582 & $+$59 24 56.381 & 0.0108 & 11.8V & AB &  ICRF J114850.3$+$592456; NGC 3894 \\
         & 12 56 14.2339 & $+$56 52 25.237 & 0.0422 & 13.5V & AS &  ICRF J125614.2$+$565225; NVSS J125614$+$565223 \\
NGC6251  & 16 32 31.9698 & $+$82 32 16.399 & 0.0247 & 12.9V & AS &  ICRF J163231.9$+$823216; NGC 6251 \\
DA426    & 16 53 52.2166 & $+$39 45 36.608 & 0.0337 & 13.8V & AL &  ICRF J165352.2$+$394536; IERS B1652$+$398; NVSS 52$+$394536 \\
NGC6454  & 17 44 56.6070 & $+$55 42 17.161 & 0.0304 & 13.5V & G  &  ICRF J174456.6$+$554217; NGC 6454 \\
NGC5077  & 13 19 31.6696 & $-$12 39 25.074 & 0.0094 & 11.9V & G  &  ICRF J131931.6$-$123925 (VCS$-$only); NGC 5077 \\
         & 18 35 03.3896 & $+$32 41 46.856 & 0.0579 & 15.3V & AS &  ICRF J183503.3$+$324146 (VCS$-$only); JVAS $+$3241 \\
AP$-$Lib   & 15 17 41.8131 & $-$24 22 19.476 & 0.0490 & 14.8V & AL &  ICRF J151741.8$-$242219; PMN J1517$-$2422 \\
NGC6500  & 17 55 59.7823 & $+$18 20 17.669 & 0.0100 & 12.6V & G  &  ICRF J175559.7$+$182021; NGC 6500 \\
         & 01 28 08.0633 & $+$49 01 05.985 & 0.0670 & 17.2V & AS &  ICRF J012808.0$+$490105 (VCS$-$only); 87GB 5.5$+$484533 \\
         & 06 03 14.3555 & $+$06 22 27.950 &        &       & R  &  ICRF J060314.3$+$062227 (VCS$-$only); PMN $+$0622 \\
         & 07 02 40.4026 & $-$28 41 50.048 & 0.0073 & 12.8V & G  &  ICRF J070240.4$-$284150 (VCS$-$only); NGC 2325 \\
         & 23 27 21.9660 & $+$15 24 37.311 & 0.0457 & 16.3V & AQ &  ICRF J232721.9$+$152437 (VCS$-$only); 2MASX 2195$+$1524375 \\
         & 23 47 04.8366 & $+$51 42 17.881 & 0.0440 & 15.5V & AL &  ICRF J234704.8$+$514217 (VCS$-$only); 2MASX 0479$+$5142179 \\
IIIZW2   & 00 10 31.0059 & $+$10 58 29.504 & 0.0893 & 15.0V & AS &  ICRF J001031.0$+$105829 DEF; MRK 1501 \\
         & 02 03 33.3849 & $+$72 32 53.667 & 0.3900d& 19.2V & AL &  ICRF J020333.3$+$723253 DEF; CGRaBS J0203$+$7232 \\
OQ208    & 14 07 00.3944 & $+$28 27 14.690 & 0.0766 & 15.1V & AL &  ICRF J140700.3$+$282714; NVSS J140700$+$282714 \\
NGC5675  & 14 32 39.8296 & $+$36 18 07.932 & 0.0133 & 12.7r & G  &  ICRF J143239.8$+$361807; NGC 5675 \\
         & 00 29 00.9860 & $-$01 13 41.759 & 0.0860 & 14.7R & G  &  ICRF J002900.9$-$011341 (VCS$-$only); PKS 0026$-$014 \\
         & 05 41 14.7577 & $+$55 50 43.570 &        & 14.3J & G  &  ICRF J054114.7$+$555043 (VCS$-$only); 87GB 2.2$+$554928 \\
         & 11 25 58.7419 & $+$20 05 54.337 & 0.1330 & 18.0V & G  &  ICRF J112558.7$+$200554 (VCS$-$only); 4C $+$20.25 \\
         & 13 17 39.1937 & $+$41 15 45.617 & 0.0662 & 14.7r & G  &  ICRF J131739.1$+$411545 (VCS$-$only); 87GB 6.7$+$413121 \\
         & 14 07 29.7622 & $-$27 01 04.293 & 0.0218 & 11.8R & AB &  ICRF J140729.7$-$270104 (VCS$-$only); PMN $-$2701 \\
         & 15 21 22.5436 & $+$04 20 30.135 & 0.0523 & 16.0V & G  &  ICRF J152122.5$+$042030 (VCS$-$only); JVAS $+$0420 \\
         & 15 59 01.7019 & $+$59 24 21.834 & 0.0602 & 14.3r & G  &  ICRF J155901.7$+$592421 (VCS$-$only); 87GB 4.5$+$593302 \\
         & 17 43 57.8326 & $+$19 35 09.019 & 0.0840 & 16.8V & AL &  ICRF J174357.8$+$193509 (VCS$-$only); 2MASX 5781$+$1935091 \\
         & 22 19 44.1753 & $+$21 20 53.186 & 0.2000 & 17.0V & AL &  87GB 221719.9$+$210528; source of z unclear \\
         & 01 13 43.1449 & $+$02 22 17.316 & 0.0470 & 16.0V & AL &  ICRF J011343.1$+$022217; UGC 00773 \\
         & 01 50 02.6972 & $-$07 25 48.487 & 0.0177 & 15.6V & AS &  ICRF J015002.6$-$072548; PMN J0150$-$0725 \\
UG01841  & 02 23 11.4112 & $+$42 59 31.384 & 0.0213 & 14.8V & AS &  ICRF J022311.4$+$425931; 4C $+$42.07 \\
         & 08 39 15.8276 & $+$28 50 38.803 & 0.0791 & 14.9V & G  &  ICRF J083915.8$+$285038; B2 0836$+$29 \\
         & 15 16 40.2190 & $+$00 15 01.908 & 0.0525 & 16.6V & AB &  ICRF J151640.2$+$001501; CGRaBS J1516$+$0015; NVSS 40$+$001502 \\
         & 22 04 17.6523 & $+$04 40 02.022 & 0.0270 & 15.2V & AS &  ICRF J220417.6$+$044002; 4C $+$04.77 \\
         & 23 33 55.2378 & $-$23 43 40.658 & 0.0477 & 17.0V & AB &  ICRF J233355.2$-$234340; PKS 2331$-$240 \\
         & 08 24 49.2600 & $-$24 28 52.554 &        &       & R  &  ICRF J082449.2$-$242852 (VCS$-$only); PMN J0824$-$2428 \\
 \hline
 \label{table}
 \end{tabular}
 \end{table*}

\end{document}